\documentclass[%
 reprint,
 amsmath,amssymb,
 aps,
]{revtex4-2}

\usepackage{graphicx}% Include figure files
\usepackage{dcolumn}% Align table columns on decimal point
\usepackage{bm}% bold math
\usepackage{xcolor}
\usepackage{subcaption}
\usepackage{booktabs}
\usepackage{multirow}
\usepackage{svg}
\usepackage{float}
\usepackage[export]{adjustbox}
\usepackage{flushend}

\begin{document}

\preprint{APS/123-QED}

\title{A Shift-Invariant Deep Learning Framework for Automated Analysis of XPS Spectra}% Force line breaks with \\

\author{Issa Saddiq}
\email{i.saddiq25@imperial.ac.uk}
\affiliation{%
 Department of Chemistry, University College London (UCL), London, United Kingdom
}%

\author{Yuxin Fan}
\affiliation{%
 Department of Chemistry, University College London (UCL), London, United Kingdom
}%
\author{Robert G. Palgrave}%
\affiliation{%
Department of Chemistry, University College London (UCL), London, United Kingdom
}%
\affiliation{HarwellXPS,  Research Complex at Harwell, Oxfordshire, United Kingdom}

\author{Mark A. Isaacs}%\affiliation{%
\affiliation{%
 Department of Chemistry, University College London (UCL), London, United Kingdom
}%
\affiliation{HarwellXPS,  Research Complex at Harwell, Oxfordshire, United Kingdom}

 \author{David Morgan}%\affiliation{%
\affiliation{%
Department of Chemistry, Cardiff University, Cardiff, United Kingdom
}%
\affiliation{HarwellXPS,  Research Complex at Harwell, Oxfordshire, United Kingdom}

\author{Keith T. Butler}%
\email{k.t.butler@ucl.ac.uk}
\affiliation{%
Department of Chemistry, University College London (UCL), London, United Kingdom
}%

\begin{abstract}

X-ray Photoelectron Spectroscopy (XPS) is a crucial technique for material surface analysis, yet interpreting its spectra is often challenging for both human analysts and automated methods due to the prevalence of variable spectral shifts and overlapping peaks.  This project introduces a machine learning solution using a Spatial Transformer Network (STN), a type of neural network that implicitly learns to align spectra. An STN model was designed to classify the chemical environments present in an input spectrum, using functional groups as a proxy. The model was trained and tested on a large synthetic dataset of 100,000 spectra, created by linearly combining real experimental data from a library of 104 polymers. \cite{RN22} To simulate experimental variability, random uniform shifts and broadening were applied to the data. The STN was found to effectively correct for random electrostatic shifts (up to 3.0 eV) and achieved relatively high accuracy ($\sim$ 82\%) in identifying functional groups, despite utilizing a much simpler architecture than previous work. These findings demonstrate that neural networks can effectively learn the underlying relationships between spectral features and chemical composition when they are able to intrinsically account for variable shifts. This work advances the development of more reliable automated XPS analysis, offering potential as an assistive tool for researchers and as a core component in future autonomous systems like self-driving laboratories.

\end{abstract}

\maketitle

%\tableofcontents

\section{\label{sec:level1}Introduction}

X-ray Photoelectron Spectroscopy (XPS) is an indispensable tool for surface analysis across a wide variety of fields, including nanoscience, metallurgy, semiconductors, organic materials and electrochemistry.\cite{RN1}  However, extracting meaningful chemical information from XPS spectra typically requires expert interpretation. The number of experts in this field has not increased proportionately with the increasing popularity of XPS techniques.\cite{RN9} Today, most XPS is performed by non-surface scientists who publish their work in non-surface journals.\cite{RN4} This expertise gap, coupled with growing data volumes from modern high-throughput XPS workflows,\cite{RN3} creates a bottleneck and increases the risk of data misinterpretation,\cite{RN2} highlighting the urgent need for more robust and reliable automated analysis methods.

Automating the nuanced process of expert XPS analysis presents significant challenges. The inherent variability within experimental spectra makes it difficult to create robust, rule-based programs that rely solely on predefined features or intensity thresholds in specific spectral regions. Notable challenges for automated analysis are overlapping peaks from different chemical environments and unpredictable binding energy (BE) shifts induced by sample surface charging. Machine Learning (ML) offers a promising approach to overcome these difficulties in spectral complexity and variability, due to its ability to automatically learn and identify complex patterns, subtle peak shifts, and correlations directly from large amounts of spectral data, bypassing the need for explicitly programmed rules. Previous studies have already demonstrated the feasibility of ML models (specifically neural networks (NNs)) for tasks like quantification of atoms/environments from XPS spectra.\cite{RN5,RN6}

More generally, there have been numerous studies in recent years developing machine learning (ML) approaches to facilitate the analysis of spectroscopic data. These studies have encompassed a diverse range of techniques, from X-ray absorption and emission to IR/Raman and inelastic neutron scattering.\cite{penfold2024machine, penfold2023deep, chen2024robust, zheng2018automated, chen2021machine, butler2021interpretable, liu2017deep, poppe2025autoencoding} Acceleration of characterisation techniques is a key step in the development of autonomous laboratories, and there have been several high-profile demonstrations of the incorporation of high-throughput characterisation in materials discovery workflows.\cite{szymanski2023autonomous, hung2024autonomous}

However, automated analysis can sometimes fail to account for important experimental artefacts.\cite{leeman2024challenges, RN12, butler2021interpretable} One of the major challenges in training ML models for characterisation is the lack of large, reliable labelled datasets. This issue is often addressed by using simulated data to train supervised ML models, which are then applied to analyse experimental data.\cite{chen2021database, mathew2018high} Yet, the well-documented phenomenon of domain shift, where the application case contains features not represented in the training distribution, means that models developed in this manner are often brittle in real-world applications.\cite{kouw2018introduction}

Several studies have proposed methods for mitigating domain shift, including the use of generative machine learning.\cite{morgan2020opportunities, anker2023using, hu2024realistic} In this study, we address a common source of domain shift, i.e. arbitrary uniform shifts of spectra arising from sample and instrument effects. While our investigation focuses on XPS, the approach we demonstrate is lightweight and should be straightforward to incorporate into any ML-based spectroscopy analysis method.

Variable energy shifts, primarily caused by an experimental effect called surface charging, apply a uniform offset to the entire spectrum\cite{RN21}. Conventional NNs struggle to account for this translational variance, interpreting two chemically identical spectra with slightly different energy shifts as entirely separate inputs. This limitation is borne out in practice; when a conventional NN is trained and tested on spectra with randomized uniform shifts applied (where a constant offset is applied to the entire energy range of a single spectrum, but the magnitude of this offset varies randomly between different spectra), there is a significant drop-off in prediction accuracy. This highlights the need for architectures that can explicitly and efficiently handle such variations.

To solve this problem, this work introduces a novel one-dimensional Spatial Transformer Network (STN) architecture. The STN component learns to spatially transform the input by calculating the optimal alignment for a given spectrum before it is passed to the conventional classification layers. We train the model on a large dataset of polymer XPS spectra, with the objective of inferring which functional groups are present in the spectrum. The dataset is comprised of 104 experimental spectra, which are then mixed in varying ratios and under a mixture of transformations (see methods for more details) to create a ``synthetic'' dataset of 100,000 labelled spectra.

 When tested on synthetic spectra with random uniform shifts applied (up to 3eV), the STN model achieved a prediction accuracy of $\sim$82\%. This stands in stark contrast to the much lower performance of alternative models of equivalent complexity, such as a standard fully connected neural network (Multilayer Perceptron, MLP) and a Convolutional NN (CNN), which both performed at $<$55\%. Further analysis reveals that this superior performance stems from the STN's ability to resolve fine-structure details, allowing it to correctly distinguish between distinct FGs with peaks at similar binding energies where conventional NNs typically fail.

\section{\label{sec:level2}Background}

\subsection{Neural Networks (NNs)}

To understand why variable shifts are so detrimental, we must first consider how a standard neural network processes spectral data. The model's objective is to learn the relationship between input features and output labels. In practice, it takes a spectrum as a one-dimensional input tensor (a list of intensities at fixed binding energies) and processes it through interconnected layers of nodes. The final layer produces an output tensor predicting the presence or absence of each environment (represented as FGs in this work). During training, the network's internal parameters are iteratively adjusted to minimise the error between its predictions and the true labels for a given spectrum.

This architecture is fundamentally vulnerable to sample charging, a known challenge in XPS interpretation where a surface potential, usually of unknown magnitude, develops on the sample during analysis. This can result in a near-uniform shift of all spectral features of the charging species to higher or lower binding energy.\cite{RN9} Because each input node is tied to a fixed binding energy, such a shift activates a completely different set of nodes, making the spectrum appear as an entirely new sample to the network.

\begin{figure}[ht]
    \centering
    \includegraphics[width=\columnwidth]{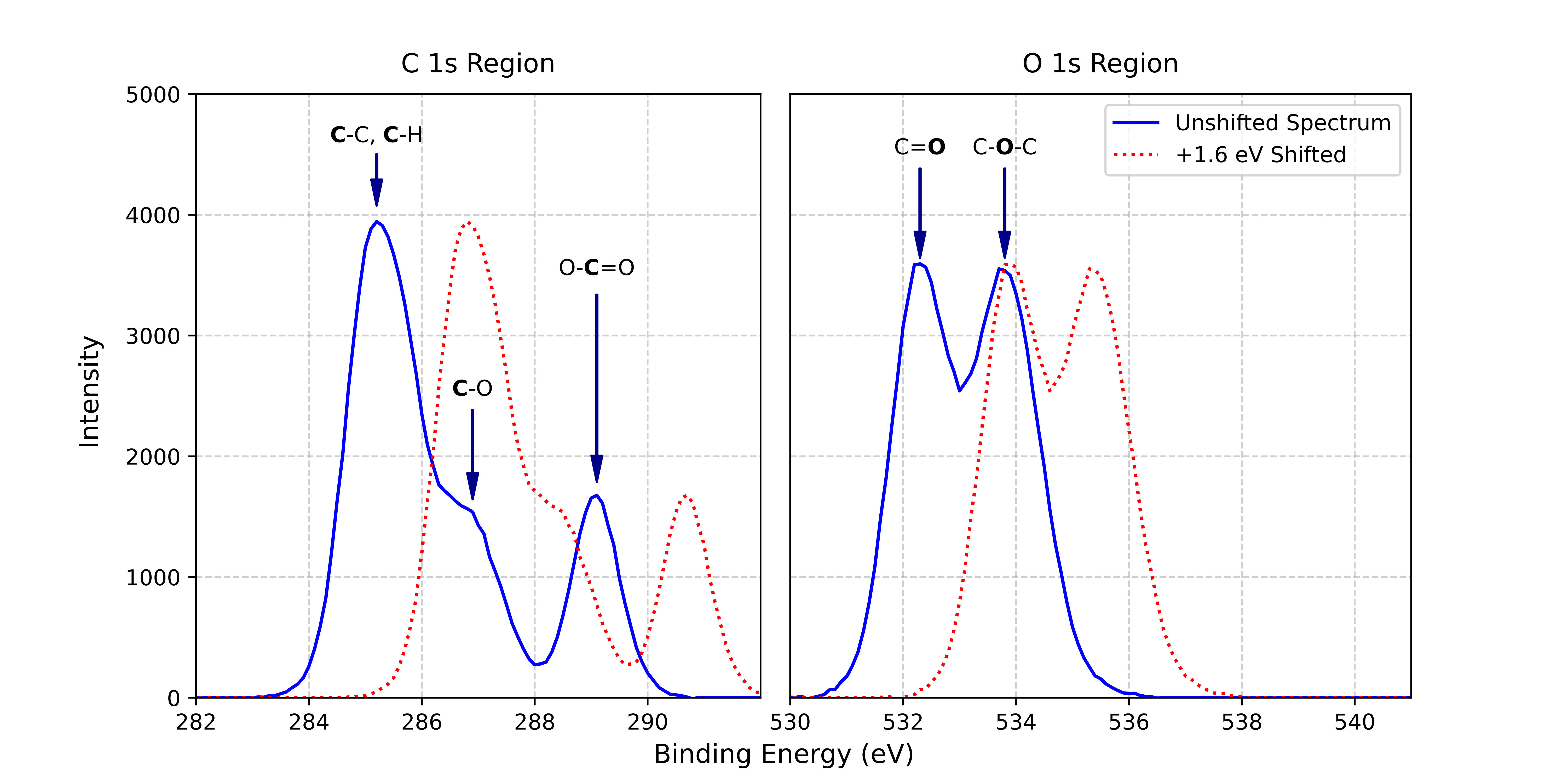}
    \caption{High-resolution C 1s and O 1s core level XPS spectra of poly(methyl methacrylate) (PMMA).\cite{RN22} Peak assignments for the primary chemical environments are annotated on the experimental data (blue line). A duplicate spectrum with a uniform +1.6 eV shift (red, dotted line) is overlaid for comparison.}
    \label{fig:shift_effect}
\end{figure}

To compensate, the model is forced to learn an inefficient  ``many-to-one'' mapping, associating a wide range of possible binding energies with a single chemical environment. This approach fails when the electrostatic shift is comparable to, or larger than, the subtle chemical shifts that distinguish different functional groups.

Figure \ref{fig:shift_effect} illustrates this critical ambiguity. The +1.6 eV shift in the PMMA spectrum (red, dotted line) causes the C=O peak to align with the original binding energy of the C-O-C peak. This overlap makes the absolute binding energy of an individual peak an unreliable feature, as it becomes impossible to disentangle the electrostatic and chemical shift contributions. In such cases, a conventional NN loses the ability to resolve fine spectral details and is forced to make ambiguous predictions.

\subsection{Convolutional Neural Networks (CNNs)}
Previous work has employed a specific type of NN called Convolutional Neural Networks (CNNs), which are powerful tools for pattern recognition.\cite{RN5, RN6} CNNs operate by sliding learned filters, or kernels, across the input data to identify local features, such as the characteristic shape of a peak. While this mechanism has proven effective for tasks like noise reduction in XPS analysis, its rationale for achieving shift invariance is less clear.\cite{RN6} The primary features identifying chemical states in XPS are the precise peak positions (chemical shifts), not necessarily their shapes or widths. Therefore, while CNNs offer a degree of spatial robustness, it is not well established how their feature extraction process directly addresses the problem of large, variable electrostatic shifts, and no systematic comparison has been performed in this context.

\subsection{\label{sec:STN_explain}Spatial Transformer Networks (STNs)}

This work introduces a Spatial Transformer Network (STN) to more explicitly handle the challenge of variable shifts in XPS data.  Originally developed for image recognition\cite{RN28}, the STN is a module added to the start of a neural network that learns to dynamically align each input spectrum. By analysing the entire spectrum at once, the STN can leverage the consistent relative positions between various peaks to infer the global electrostatic shift. This allows the model to achieve shift invariance by focusing on the stable relationships between features, rather than relying on their absolute and unreliable binding energy positions.

This alignment process is learned implicitly as an integrated part of the model's training loop. The STN is not explicitly instructed to align spectra to a known reference peak. Instead, the module is trained end-to-end with the main classification network, learning to apply whatever spatial transformation will best standardise the input to maximise the final prediction accuracy.  It achieves this by mapping all shifted variations of a given spectrum to a consistent, internally learned \textit{canonical representation}. This flexible approach provides a more transparent mechanism for achieving shift invariance than a CNN. The STN learns to perform a global alignment by leveraging the full spectral context—using the learned relationships between disparate peaks (such as related C 1s and O 1s features both represented in a C-O group) to infer the correct shift, rather than relying on a single reference point or abstract local patterns. In this way, the STN's holistic approach of using the entire spectral context to inform alignment is hypothesised to approximate the strategy of a human expert.

\begin{figure}[ht]
    \centering
    \includegraphics[width=\columnwidth, trim={0pt 120pt 0pt 0pt},
        clip]{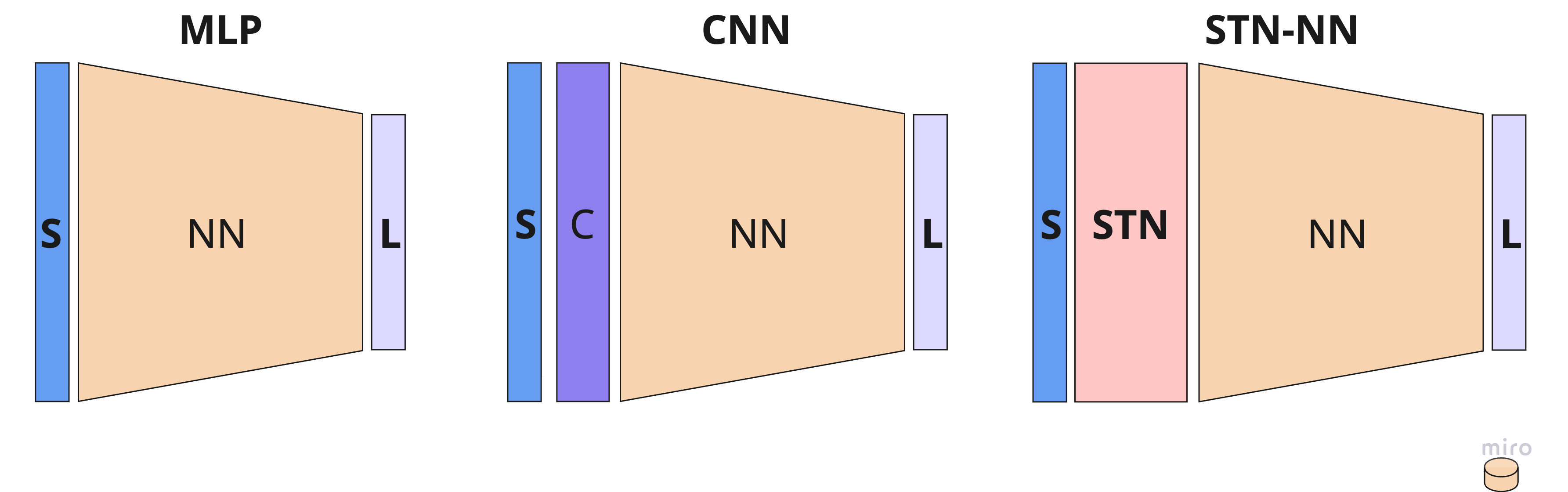}
    \caption{Diagram comparing architectures of a Multilayer Perceptron (MLP), a Convolutional Neural Network (CNN) and a Spatial Transformer (STN)-based NN, and how they process a spectrum (\textbf{S}) into predicted label (\textbf{L}).}
    \label{fig:classifier_diagrams}
\end{figure}

\section{\label{sec:level3}Methods}

This study focuses on creating models for chemical environment detection rather than full quantification, with the view that the techniques developed to achieve shift invariance can be implemented in more complex quantification algorithms in future work. By treating the problem as one of classification, this approach isolates variable shifts as a source of error, providing a direct measure of how well a given architecture can resolve peak-environment relationships.

To train and test such classification models, it was first necessary to create a large dataset of XPS spectra, each with a corresponding label representing the chemical environments present. Unfortunately, public datasets of the required size for training NNs are not currently available for this task. Instead, this work utilized existing experimental spectra to generate a large dataset of synthetic spectra, following the methodology established by Pielsticker et al.\cite{RN6}

\subsection{Synthetic Spectra Generation}

The training data was taken from the Scienta300 ESCA Polymer Database. \cite{RN22}  From this database of 111 experimental polymer spectra recorded using a high-resolution monochromated spectrometer, 104 were selected for synthetic data generation. Polymers with unknown compositions (e.g., ambiguous copolymer blends) were excluded. The database includes peak fitting by the authors, but this was not used here. We used the raw data and carried out background subtraction using a Shirley background for each core line region. The binding energy limits of the background were chosen on a case-by-case basis to remove the inelastic background to the best of our judgement. The valence region was not included in the training data because it is less informative for identifying functional groups: valence peaks are relatively broader and more closely spaced than core peaks, leading to significant overlap and ambiguity in peak assignment \cite{RN26}. \textcolor{black}{Therefore, each core line region was concatenated into a single input vector representing the full core level range (40–700 eV).}

To frame this as a multi-label classification problem, each base spectrum was assigned a binary label vector $(e_i)_j$ of length $n$ where $n$ is the number of independent labels in the dataset. 
\begin{equation}
    (\mathbf{e}_i)_j = 
\begin{cases}
1 & \text{if } i \text{ present}, \\
0 & \text{if } i \text{ absent},
\end{cases}
\quad j = 1, \dots, n.
\end{equation}
Each element in this vector ($i$) corresponds to a predefined functional group (FG), \textcolor{black}{indicating its presence (($\mathbf{e}_i)_j = 1$) or absence (($\mathbf{e}_i)_j = 0$)} within the polymer's repeat unit. In this system, each FG class effectively represents all of the distinct constituent chemical environments that make up that FG.

To create a diverse training set, new spectra were generated by taking random linear combinations of the 111 processed and labeled base spectra, simulating a wide range of realistic polymer compositions. To ensure the models would be robust to real-world experimental variability, two critical artifacts were subsequently simulated.

\textcolor{black}{First, to simulate instrumental effects and experimental variance in spectral resolution, random Gaussian broadening was applied to each synthetic spectrum. The broadening width, $\sigma$, for each sample was drawn from a half-normal distribution with a standard deviation of 0.2 eV. This specific distribution ensures that 95\% of the generated broadening parameters remain below 0.43 eV, effectively capturing realistic instrumental broadening while remaining below the empirical limit of $\sigma$  = 0.5 eV. As illustrated in \ref{fig:broadening}, exceeding this limit causes adjacent chemical environments (such as the alkane (C-C) at ~285.0 eV and alcohol (C-OH) at ~286.5 eV) to become spectrally indistinguishable, representing a loss of chemical information beyond what can reasonably be resolved.}

Second, a random uniform energy shift was applied over a range of up to ±5 eV to mimic the unpredictable effects of sample surface charging. This range exceeds the ±3 eV shifts simulated in previous work, allowing the model's performance to be tested beyond typical experimental variability.\cite{RN6}

\textcolor{black}{To eliminate intensity bias and ensure the model remains invariant to non-chemical experimental factors, all synthetic spectra were area-normalized such that the total integral over the 40–700 eV range equals unity. This normalization improves training stability and model generalisability by forcing the network to prioritize the learning of relative peak profiles and ratios, rather than absolute intensity counts that are sensitive to the acquisition parameters and signal variability of the original experimental dataset.}

Polymers serve as an ideal test case for this method, as their structures contain numerous, distinct chemical environments of the same elements (e.g., C 1s in C-C, C-O, C=O), providing a sensitive benchmark for a model's ability to resolve subtle chemical state information.

\subsection{\label{sec:STN_implementation}STN Implementation}

To implement the STN, this work leverages PyTorch's affine transformation functions, which are designed for 2D images.\cite{RN28} To apply these to 1D data, each spectrum is treated as a pseudo-2D image of size $1 \times L$.

This requires the use of a $2 \times 3$ affine transformation matrix, $\theta$. For a pure horizontal shift, this matrix is defined as:
\begin{equation}
    \theta = \begin{bmatrix} 1 & 0 & t \\ 0 & 1 & 0 \end{bmatrix}
    \label{eq:affine_matrix_shift_only}
\end{equation}
Here, the network's sole task is to learn the optimal translation (shift) parameter, \textbf{$t$}, which occupies the position in the matrix corresponding to the translation along the x-axis. The scaling factor is fixed to 1, and the second row ensures the transformation has no effect on the vertical pseudo-dimension.

The transformation is applied using an inverse mapping strategy. For each point $x'_{out}$ in the target (aligned) grid, the STN computes the corresponding source coordinate $x_{src}$ in the original input spectrum. The full matrix multiplication simplifies to the direct 1D transformation:
\begin{equation}
    x_{src} = x'_{out} + t
    \label{eq:stn_1d_map_shift_only}
\end{equation}
In practice, PyTorch functions (\texttt{F.affine\_grid} and \texttt{F.grid\_sample}) use the learned $\theta$ matrix to compute this mapping and generate the final, aligned spectrum, using linear interpolation to sample intensities at non-integer source coordinates.

\textcolor{black}{One notable advantage of the STN's modular approach is its simplicity; the STN localization network adds a negligible number of trainable parameters to the overall architecture. By outputting a single translation parameter ($t$) for the transformation, the STN enables shift invariance without the high computational or parametric cost associated with deep convolutional or recurrent layers.}

\subsection{Baseline comparison}

This synthetic dataset was used to train and test three distinct architectures for comparison: a Multi-layer Perceptron (MLP), a Convolutional Neural Network (CNN), and the proposed STN-based NN. All models were trained for the classification task using a Binary Cross Entropy loss function, with precise architectural and training details presented in the appendix \ref{sec:architectures}.

\textcolor{black}{By utilizing the same MLP classifier as the foundational `backbone' for all three architectures, we ensure that the performance differences observed are directly attributable to the specific input-processing mechanisms (convolutional kernels vs. spatial transformation) rather than variations in total model capacity.}

Final model accuracy was then evaluated on this same test set, calculated as the percentage of samples where the predicted binary label vector exactly matched the corresponding ground-truth vector ('exact match accuracy'). This metric offers a practical interpretation compared to the loss function used during training. Per-class performance was further analyzed using confusion matrices  to diagnose error patterns like false positives (overprediction) and false negatives (underprediction) for each functional group.

\textcolor{black}{Hyperparameters, such as the number of nodes per layer and dropout rates, were optimized for the baseline MLP to maximize predictive accuracy while minimizing the risk of overfitting. This stability was verified through the inspection of training and validation loss curves (see SI), ensuring the test loss remained converged. Following this optimization, the STN and CNN modules were integrated using this same framework to isolate the impact of their respective mechanisms.}
\section{\label{sec:level4}Results}

\subsection{Performance Comparison}
\begin{figure}[ht]
    \centering
    \includegraphics[width=\columnwidth]{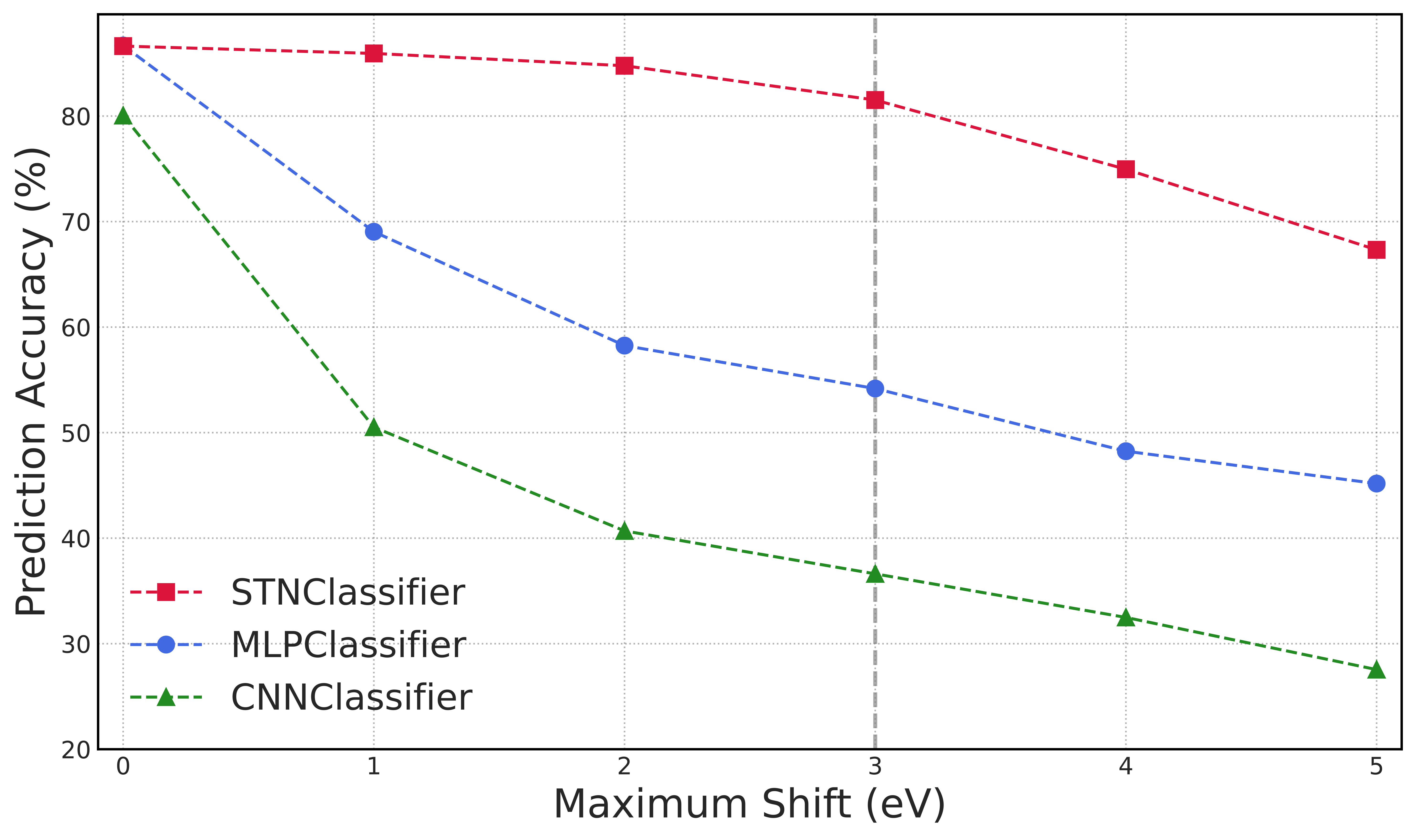}
    \caption{Model test classification prediction accuracies for models (multilayer Perceptron (MLP), Spatial Transformer Network-Neural Network (STN-NN) and convolutional neural network (CNN, kernel size = 3.0eV)) trained and tested at different levels of maximum shift.}
    \label{fig:classifier_comparison}
\end{figure}

Figure~\ref{fig:classifier_comparison} displays the classification accuracy versus maximum spectral shift for the three models: a multilayer perceptron (MLP), a Convolutional NN (CNN), and an STN-enhanced NN (STN-NN). The MLP (basic NN) shows a distinct decline in accuracy as shift increases. \textcolor{black}{This performance degradation is understandable given the NN's approach of treating spectral points independently. Increasing the shift effectively 'blurs' features by averaging peaks over a wider range during training and testing. Consequently, the NN learns broader, less defined 'active' regions for functional groups (FGs). While this may lower detection thresholds, it also causes a loss of fine spectral details and increases overlap between FG regions, ultimately reducing prediction confidence and accuracy.}

The initial hypothesis that a single convolutional layer could inherently account for small spectral shifts appears contradicted by the observed performance trend of the CNN. Any potential benefit from the kernel's local translation invariance seems outweighed by the competing effect of the convolution distorting or losing fine spectral details, ultimately leading to worse performance than the MLP at higher shift levels.
Additionally, the added complexity of the convolutional layer may hinder training, for example by encouraging overfitting to local patterns in the training data.

As hypothesized, the Spatial Transformer Network (STN) proved highly effective at enabling the downstream neural network to account for spectral shifts. This approach maintained a very strong classification accuracy relative to the baseline MLP and the alternate CNN model. At a maximum shift of 3.0 eV (the same augmentation employed by Pielsticker et al.\cite{RN6}), the STN's accuracy dropped by only 5\%.  This demonstrates a significant improvement in robustness compared to the baseline MLP and CNN models, whose accuracies fell by 32\% and $\sim$50\% respectively. This performance confirms the STN's capability to effectively standardise the input spectra before classification. The broader implications of these encouraging results, alongside a critical examination of potential overfitting to patterns within the synthetic datasets, will be addressed in detail in the Discussion section.

While this doesn't rule out the potential success of a more carefully designed CNN for polymer XPS analysis, these findings suggest such complex models are prone to overfitting in this context and highlight the strengths of the simpler STN-based approach, which yielded more robust and reliable results in this study.

\subsection{Further Analysis}

% --- Start of the updated table ---
\begin{table}[htbp] % 1. Use the 'table' float environment
    % 2. Place caption and label inside the float
    \caption{Row-normalized confusion matrices (showing recall) for MLP, CNN, and STN-NN models at a 3 eV maximum shift. Results are aggregated across all FG classes. Labels AP, AN, PP, and PN denote actual positive, actual negative, predicted positive, and predicted negative, respectively.}
    \label{tab:confusion_matrices}
    
    % 3. Wrap the tabular content in 'ruledtabular' for correct line styling
     \begin{ruledtabular}
        \begin{tabular}{ccc}
        
        \textbf{CNN} & \textbf{MLP} & \textbf{STN-NN} \\ 
        \hline
        
        \includegraphics[width=0.3\columnwidth, keepaspectratio]{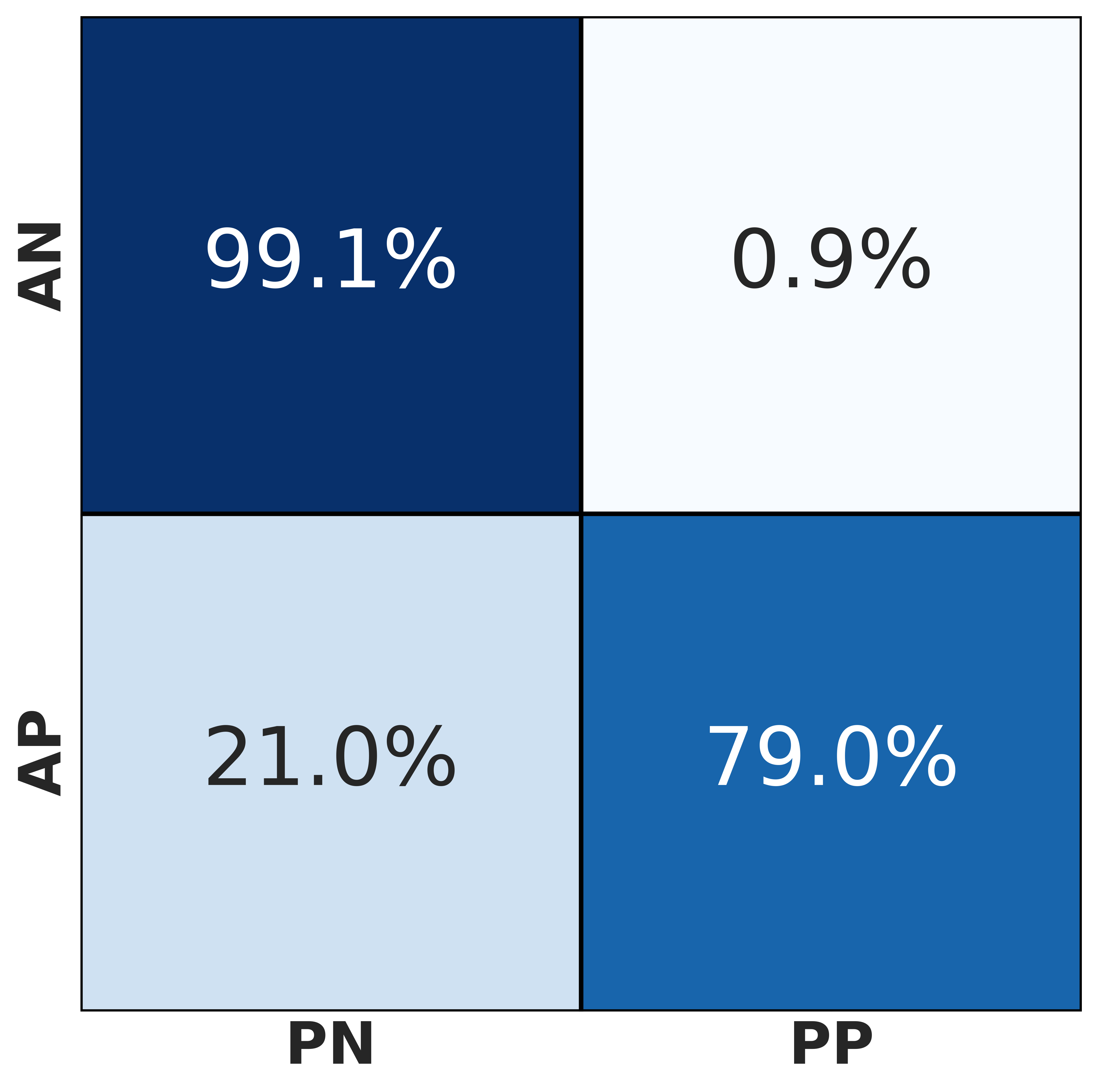}
        &
        \includegraphics[width=0.3\columnwidth, keepaspectratio]{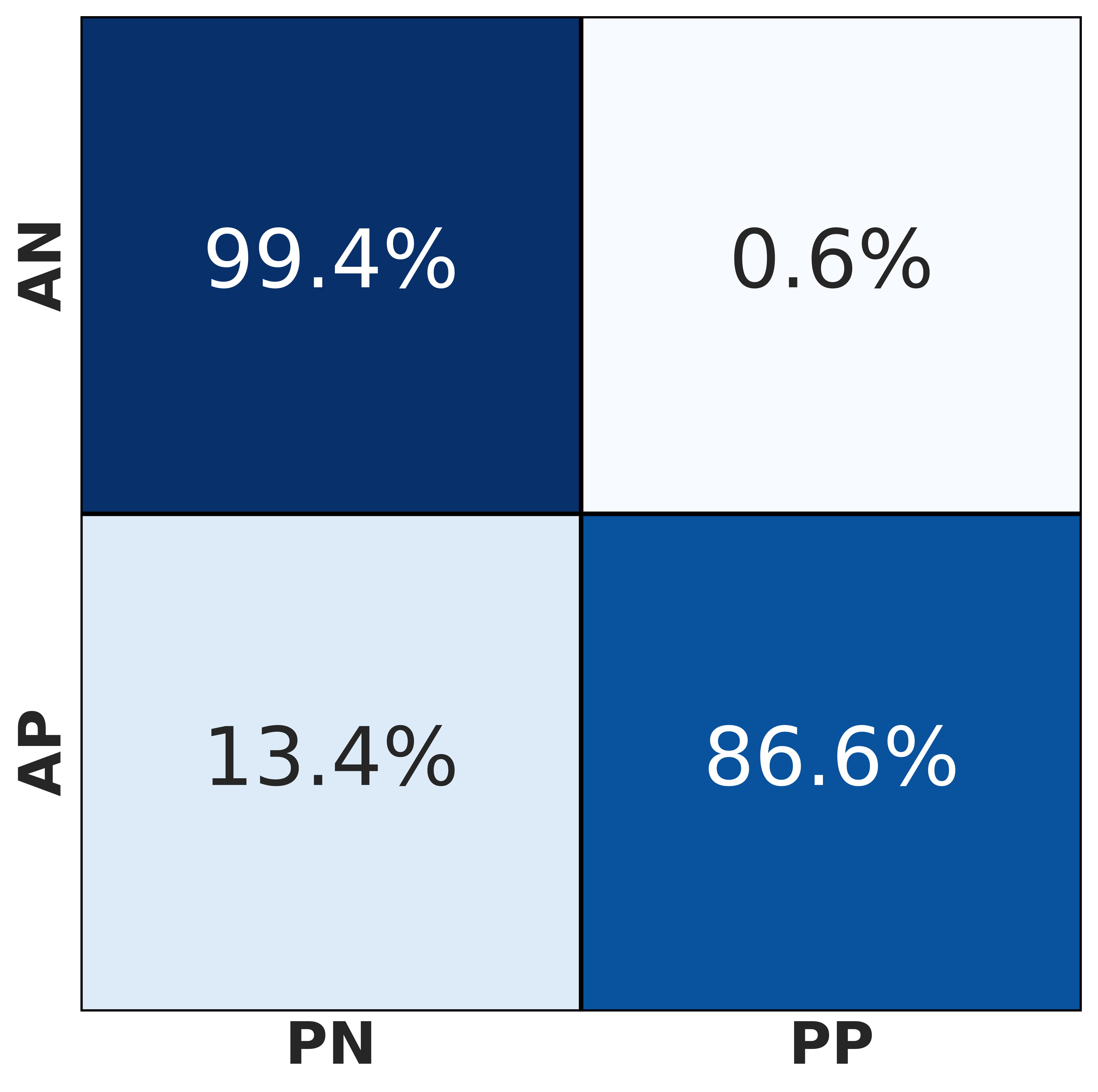}
        &
        \includegraphics[width=0.3\columnwidth, keepaspectratio]{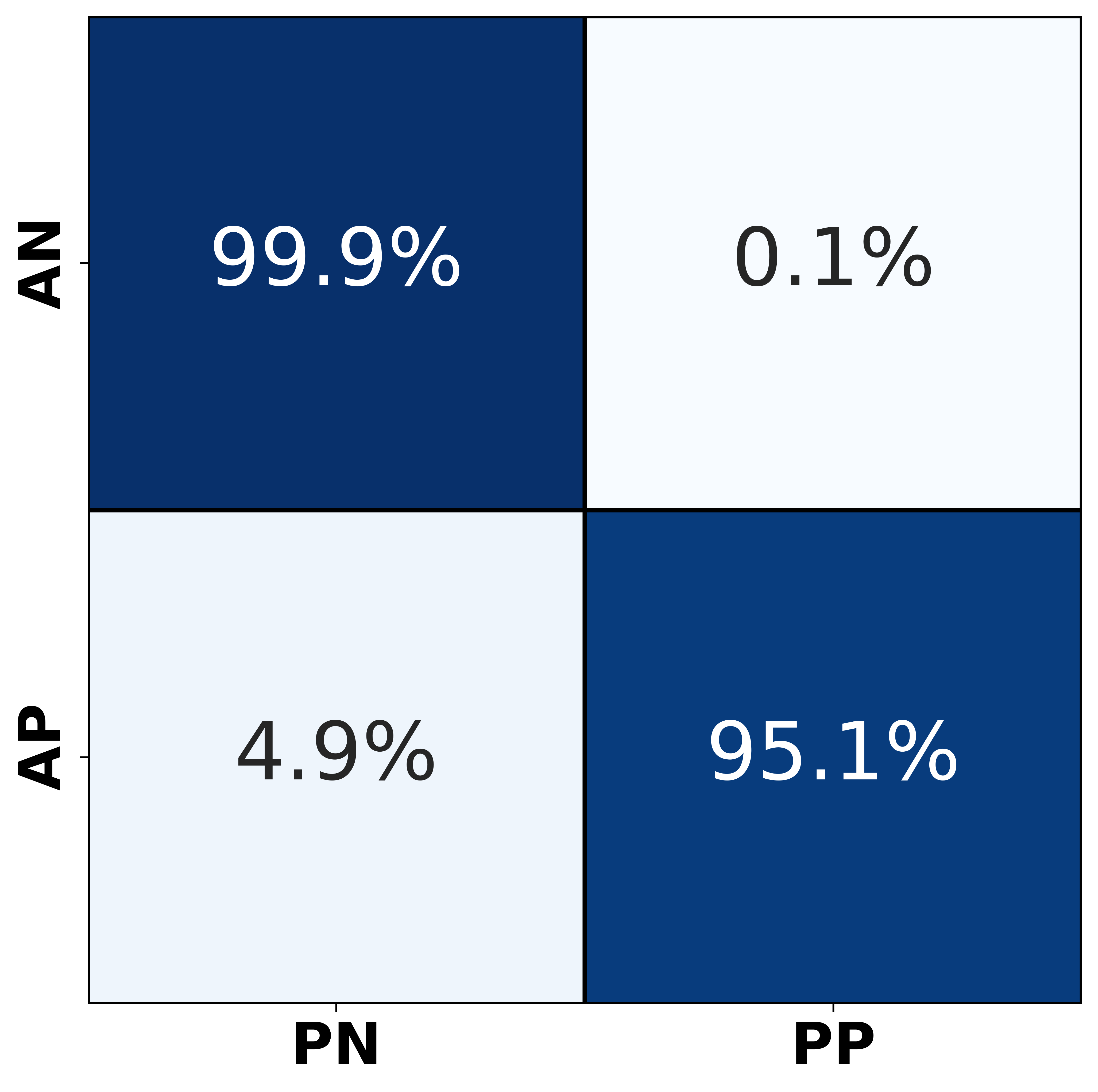} \\
        
        \end{tabular}
    \end{ruledtabular}
\end{table}
The aggregate confusion matrices shown in Table~\ref{tab:confusion_matrices} reveal the primary source of error for all models was false negatives (FN), i.e., failing to detect present functional groups. This issue is directly measured by sensitivity (the true positive (TP) rate), which quantifies a model’s ability to identify positive cases (Sensitivity = TP / (TP + FN)) and proved to be the key differentiator between model performances.

In contrast, precision, which measures the reliability of a positive prediction (Precision = TP / (TP + FP)), was consistently high for all models ($>$99\%), indicating that they rarely made false positive errors. This trade-off between lower sensitivity and higher precision is a classic outcome of training on sparse data. Because the vast majority of functional groups are absent in any given sample, the models develop a negative bias, becoming hesitant to predict the rare "present" class. This conservatism minimizes false positives but consequently increases the number of false negatives.

With this in mind, investigating the classes with the lowest sensitivity reveals the primary failure modes. Table~\ref{tab:fg_sensitivity} compares the sensitivities for several challenging FG classes (Epoxide, Anhydride, Carboxylic acid, Naphthalene) across the three models trained and tested at a 3.0 eV shift \textcolor{black}{(full list of per-class sensitivities is provided in the SI (Table~\ref{tab:full_fg_sensitivity})}.

\begin{table}[htbp]
    \centering
    \caption{Sensitivities (\%) of five worst performing FG classes for the MLP, compared with sensitivities of corresponding the CNN, and the STN-NN. All models trained and tested on XPS spectra with 3eV of shift.}
    \begin{tabular}{llccc}
        \toprule
        % --- NEW SPANNING HEADER ROW ---
        % \multicolumn{<num_cols>}{<col_spec>}{<content>}
                        & \multicolumn{3}{c}{Sensitivity   \%} &  \\
        \textbf{FG}              & \textbf{MLP}   & \textbf{CNN}    & \textbf{STN-NN}   &  \\
        \midrule
        Epoxide & 5.9 & 0.0 & 63.1 \\
        Anhydride & 37.5 & 5.6 & 79.2 \\
        Alcohol (aliphatic) & 43.5 & 24.3 & 78.0 \\
        Naphthalene & 46.6 & 2.9 & 88.0 \\
        Carboxylic acid & 52.7 & 27.7 & 84.4 \\
        \bottomrule
    \end{tabular}

    \label{tab:fg_sensitivity}
    
    \smallskip
\end{table}

\begin{figure}[ht]
    \centering
    \includegraphics[width=\columnwidth]{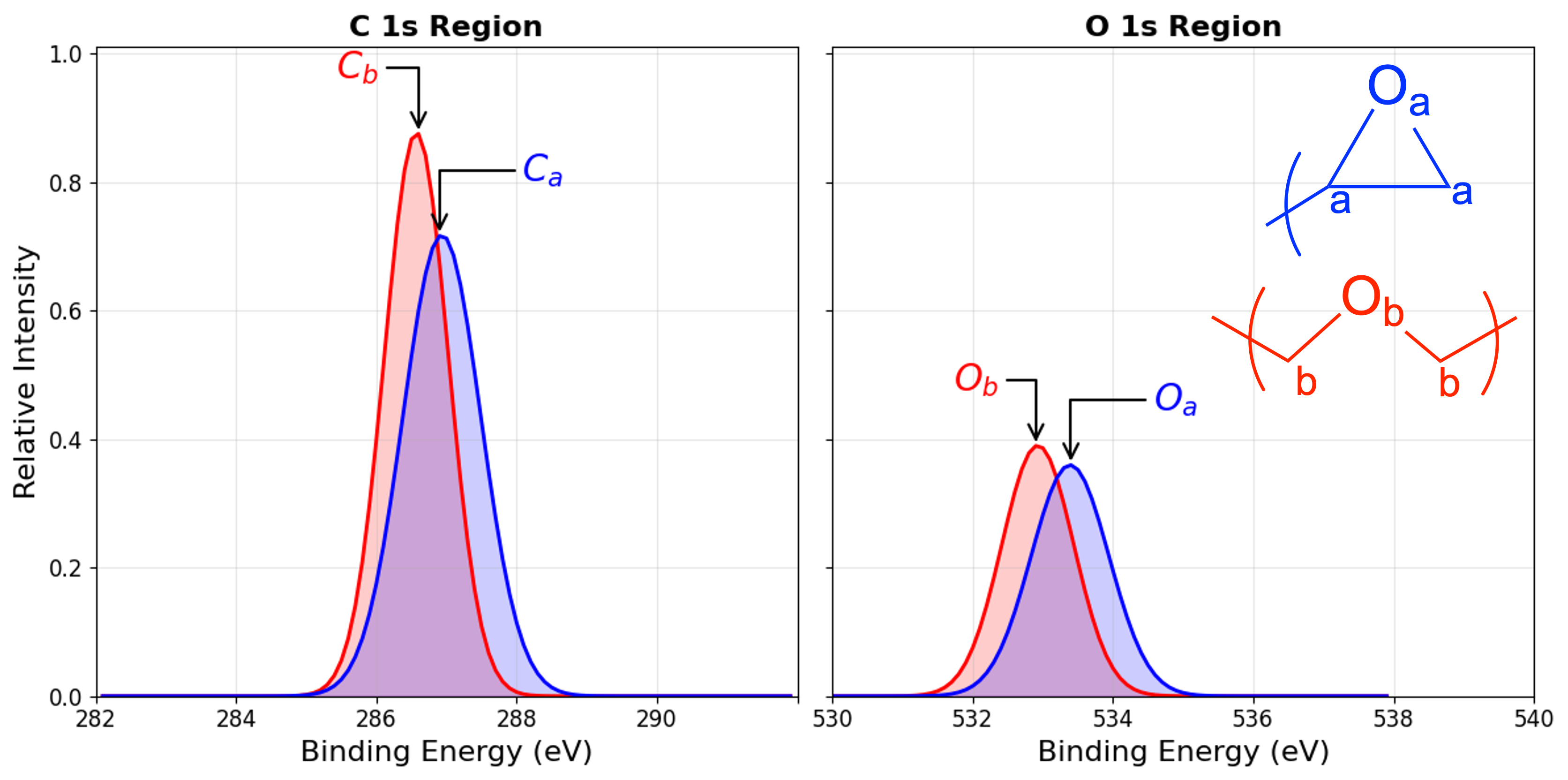}
    \caption{Comparison of C 1s and O 1s peaks for both epoxide (blue) and aliphatic ether (red) functional groups. Peaks were isolated by fitting from experimental spectra of PGMA and PEG respectively.}
    \label{fig:epoxide_comparison}
\end{figure}

\textcolor{black}{The difficulty in classifying these groups often stems from distinguishing subtle spectral variations, where slight binding energy shifts or overlapping peaks may not provide sufficiently distinct information for reliable separation. For example, distinguishing an epoxide from an aliphatic ether is highly shift-sensitive (Figure~\ref{fig:epoxide_comparison}). While their C and O environments are structurally similar, the severe bond angle strain in the epoxide ring alters the local orbital hybridization, slightly raising the binding energy of the 1s core electrons compared to a relaxed ether. However, this binding energy difference is $<$0.5~eV for both the C 1s and O 1s peaks. Such minor differences in absolute binding energies of these peaks make them difficult to resolve reliably under random experimental charging variations. Furthermore, because this small offset occurs in the same direction (both epoxide peaks appear at higher energies), the relative spacing between the peaks in both spectra is nearly identical. Given that the STN's alignment mechanism relies heavily on these relative peak positions, even the STN inherently struggles to differentiate them, achieving only 63\% sensitivity for epoxides. Distinguishing between these FGs is not a trivial matter, as accurately identifying such subtly different groups can be crucial for applications like monitoring polymer epoxidation.\cite{RN23}}

The superior performance of the STN-NN, particularly on these difficult classes, underscores how spectral shift degrades the fine details crucial for classification. The general accuracy declines with increasing shift observed in the NN and CNN models can thus be attributed primarily to the increased difficulty in distinguishing subtly different chemical environments when spectral features become blurred, especially within the complex C 1s region.

\begin{figure}[ht]
    \centering
    \includegraphics[width=\columnwidth]{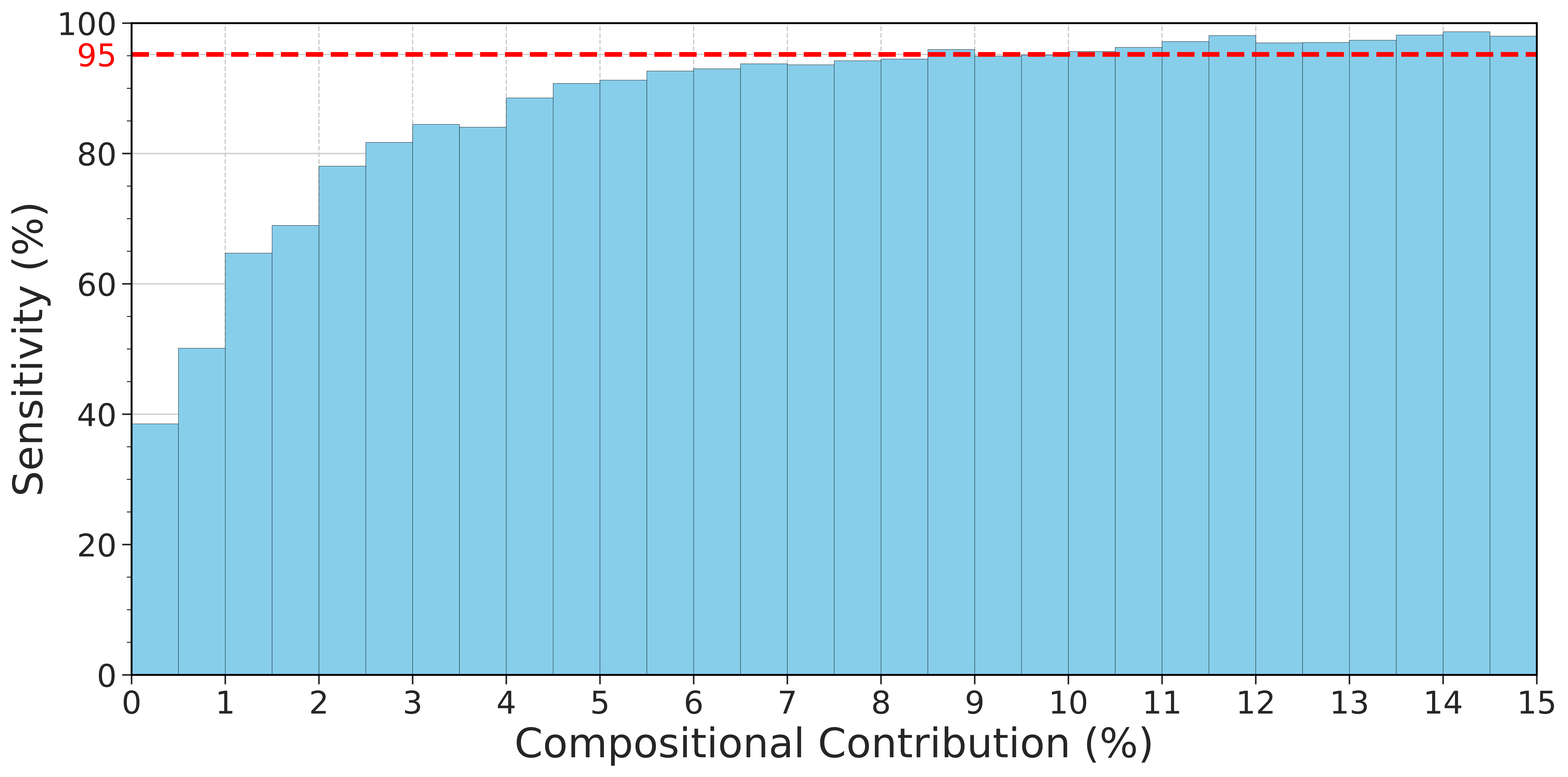}
    \caption{Histogram showing the sensitivity of the STN model as a function of compositional contribution (\%) across all functional group predictions, evaluated on spectra with a maximum 3 eV random shift applied. The horizontal red line marks the overall average sensitivity across all composition bins.}
    \label{fig:STN_detection_limits}
\end{figure}

In addition to class-level sensitivity, it is also important to consider compositional effects, that is, how detection performance varies with the concentration of a functional group in the sample and hence the prominence of its peaks. This is assessed by measuring the proportion of positive predictions within different composition ranges. Figure~\ref{fig:STN_detection_limits} shows the results for the STN model (with a maximum 3.0 eV shift applied), and the corresponding histograms for the MLP and CNN are provided in Appendix (Figure~\ref{fig:detection_limits_comparison}).

\begin{figure*}[t!]
    \centering
    % First subfigure
    \begin{subfigure}[b]{0.32\textwidth}
        \centering
        \includegraphics[width=\textwidth]{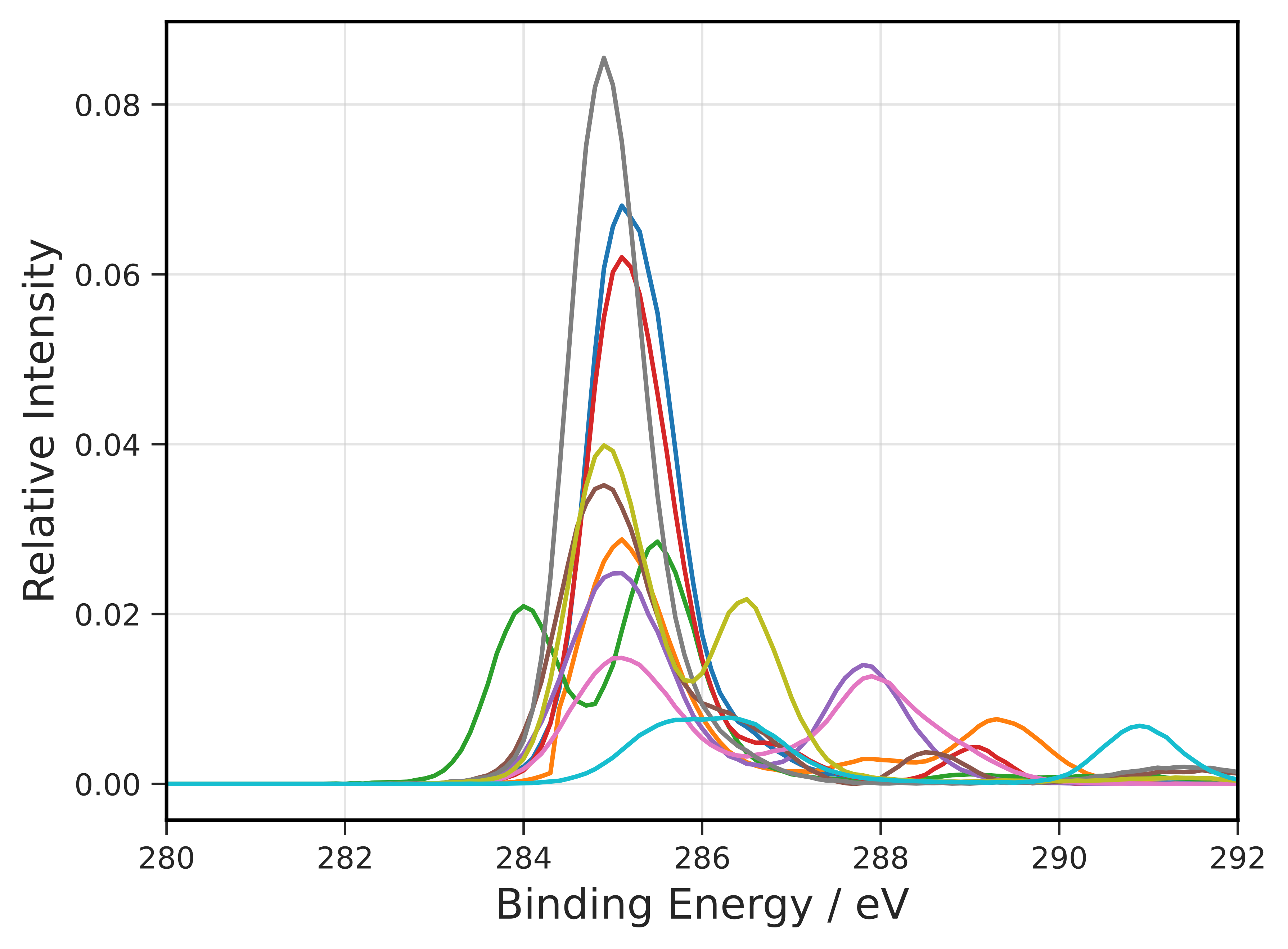}
        \caption{Input spectrum}
        \label{fig:input}
    \end{subfigure}
    \hfill % Adds horizontal space between figures
    % Second subfigure
    \begin{subfigure}[b]{0.32\textwidth}
        \centering
        \includegraphics[width=\textwidth]{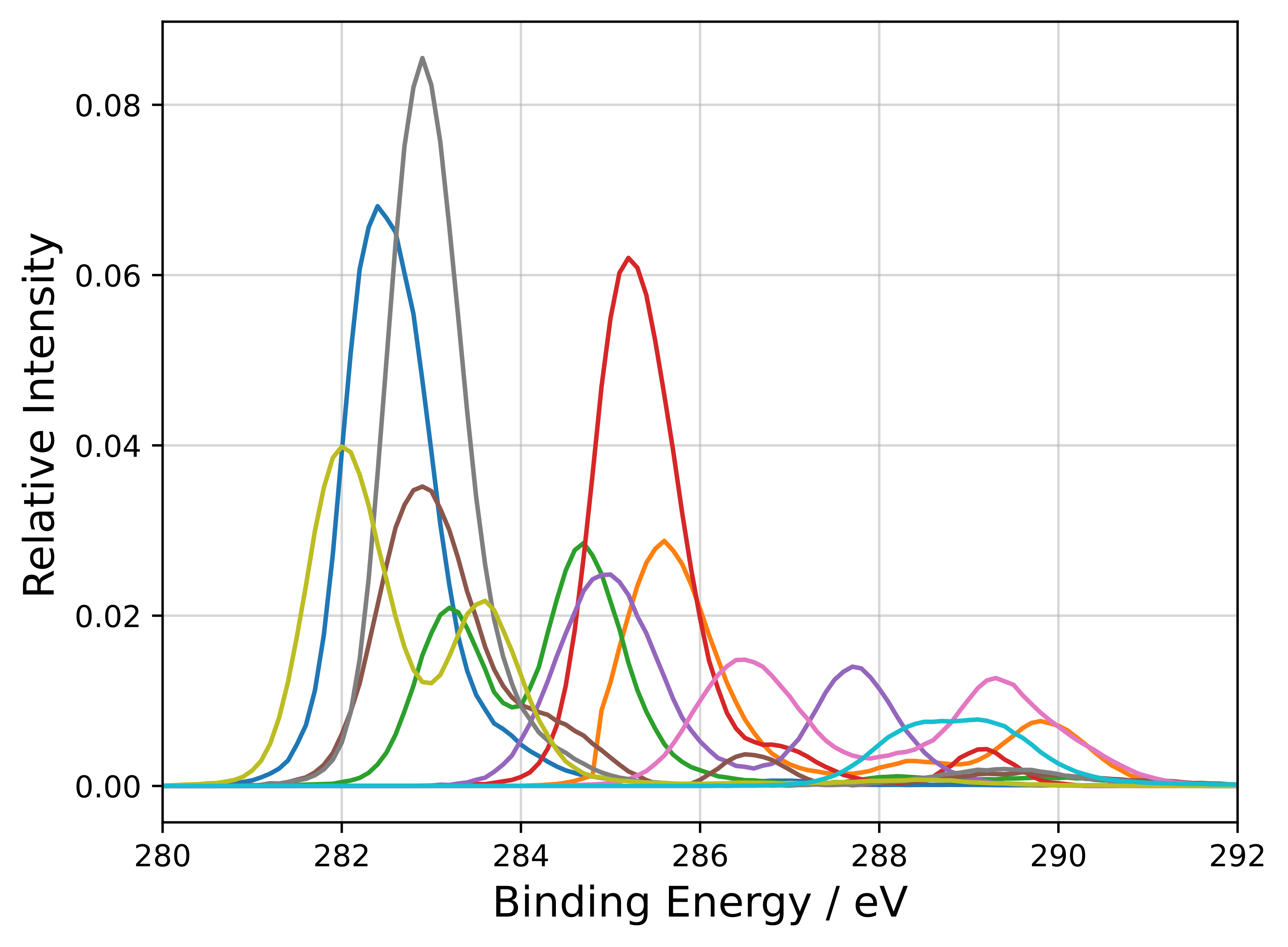}
        \caption{Shifted spectrum}
        \label{fig:shifted}
    \end{subfigure}
    \hfill % Adds horizontal space between figures
    % Third subfigure
    \begin{subfigure}[b]{0.32\textwidth}
        \centering
        \includegraphics[width=\textwidth]{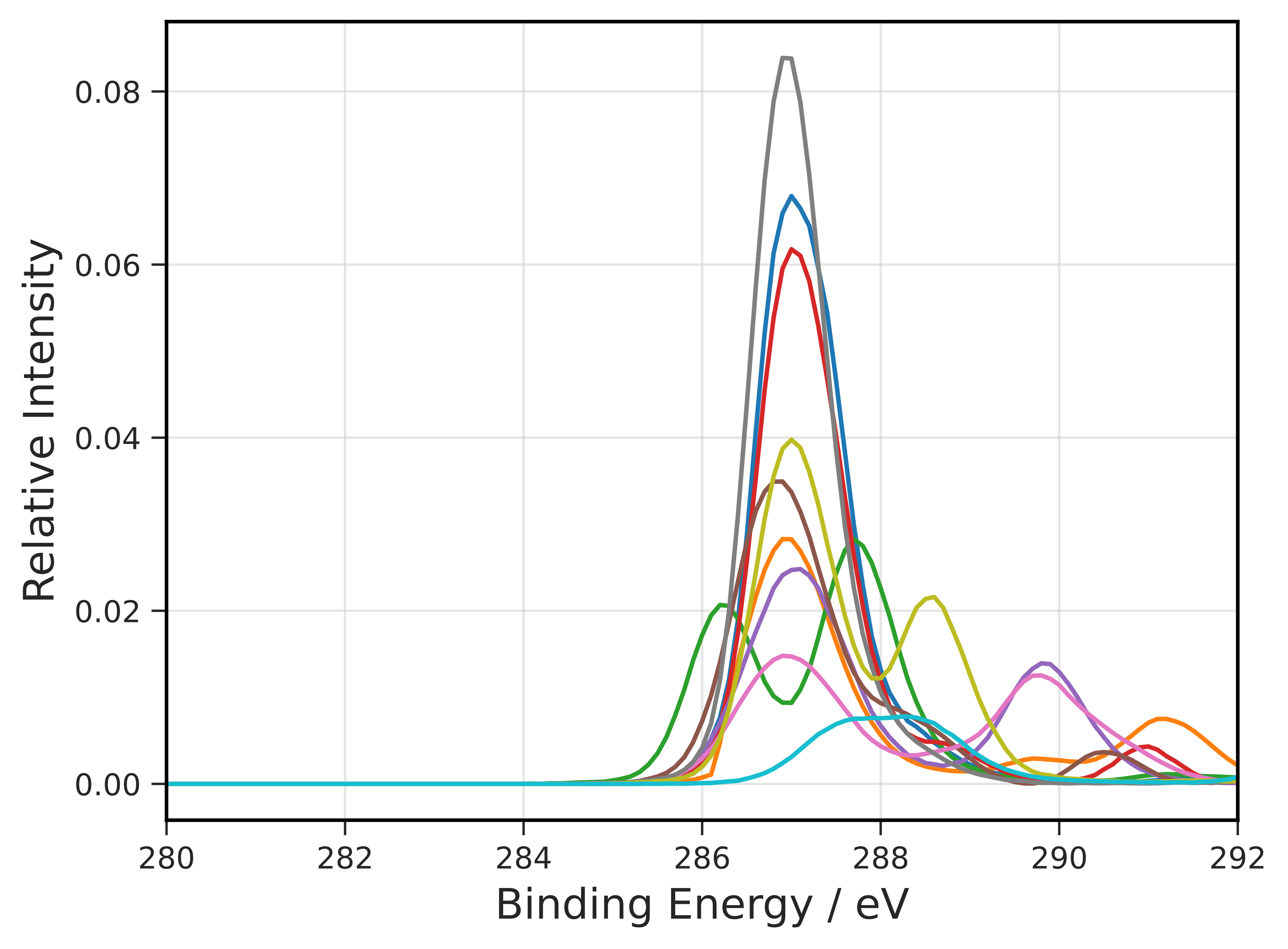}
        \caption{Aligned spectrum}
        \label{fig:aligned}
    \end{subfigure}
    
    \caption{\textcolor{black}{Process of applying random uniform shifts to a subset of 10 synthetic test spectra (a $\rightarrow$ b), and their subsequent alignment by the STN layer (b $\rightarrow$ c). All plots are cropped to highlight the C 1s region (280–292 eV).}}
    \label{fig:alignment_process}
\end{figure*}

As expected, the probability of detecting a functional group is lower at very low concentrations. The low sensitivity ($<$50\%) in the 0–1\% range is reasonable, given that the detection limit of XPS is rarely better than 0.1–1 \% \cite{XPSguide}. Sensitivity remains modest at slightly higher composition ratios (2–6\%), which is unsurprising given the high peak overlap in this dataset, where smaller peaks can be subsumed by larger ones — a challenge even for human analysts. Nevertheless, these results show the STN model performs relatively well for low concentration detection;
For comparison, the MLP is only able to achieve 87\% sensitivity at concentrations $>$12\% (Figure~\ref{fig:CNN_detection_limits}), whereas the STN is able to achieve the same sensitivity for concentrations $>$4\%.

\subsection{STN Mechanism}

Given the STN's effectiveness in handling randomly shifted spectra, it is important to investigate its internal mechanism to confirm the alignment process. This can be achieved by visualizing the STN module's output, which is the aligned spectrum that is subsequently fed into the classifier, as illustrated in Figure~\ref{fig:alignment_process}.

While the STN-aligned output spectra appear visually similar to the unshifted data, the module does not explicitly `realign' inputs to their original experimental positions, as it lacks prior knowledge of those absolute values. Instead, as mentioned in Section~\ref{sec:STN_explain}, the STN transforms inputs into a consistent, internally learned canonical representation where the absolute peak positions are arbitrary. This learned representation is effectively a local minimum that the model occupies to minimize classification loss; it lacks a hard incentive to match a physical reference as long as it can resolve the variation and relative peak spacing between different spectra.

Visual inspection of Figure~\ref{fig:alignment_process} and across the test set confirms that this requirement is met: despite all spectra being systematically aligned to different absolute binding energies, the relative spectral features remain intact. For example, the alkane peak at $\sim$285 eV consistently appears at $\sim$287 eV in the `aligned' output of the STN. This demonstrates that the STN achieves stable alignment in a chemically meaningful way, even if the absolute reference frame is shifted.

\textcolor{black}{In practical research workflows, visualizing the aligned STN output provides a valuable interpretative tool alongside raw predictions. To generate a human-interpretable spectrum for comparison with literature, the STN output can be post-processed by applying a uniform correction factor, $\Delta$. This constant shift is determined empirically by comparing the model's internal reference frame to a known test set. Restoring the spectra to standard binding energy scales allows researchers to verify the model’s internal logic and facilitates the integration of automated analysis into existing expert-led characterization pipelines.} In the above example, a $\Delta$ of $-$2 eV would restore the peaks to their expected positions. It should also be noted that small discrepancies remain even in the relative peak positions, for example, in the double-peaked spectrum shown in green in Figure~\ref{fig:alignment_process}. These residual errors explain the modest drop in classification performance, reflecting the fact that the alignment stage, while highly effective, does introduce non-negligible distortion.

\section{\label{sec:level5}Discussion}

The STN-based approach proved highly effective at achieving shift invariance for XPS classification, with performance largely independent of the shift magnitude. The effectiveness of this `pattern recognition' approach to XPS spectra can be attributed to the simplicity of both the 1D spectral data, and the well defined horizontal transformation (of varying magnitude/sign).

\textcolor{black}{While previous work by Pielsticker et al. demonstrated that CNNs can effectively handle shifted spectra, the performance difference observed here likely stems from the nature of the datasets. In transition metal spectra, chemical states are often identified by distinct, complex line shapes that a CNN can learn as shift-invariant features. In contrast, the polymer C 1s and O 1s peaks in this study share very similar profiles, making absolute and relative peak positions the primary identifiers. In this context, the added complexity of a convolutional layer appears to distort fine spectral details and encourage overfitting to local patterns, whereas the STN provides a more direct mechanism for global alignment while strictly preserving the underlying physical line shapes and relative peak separations.}

Despite the STN's strong performance, its reliability is fundamentally constrained by the synthetic training domain, a limitation shared by any machine learning approach, including previous studies that used a similar data generation process.\cite{RN6,RN5} This work builds upon that foundation by introducing the STN as a more direct and effective mechanism for handling spectral shifts. Therefore, the critical next step is to validate the model's generalizability on a diverse set of real experimental spectra to confirm its practical utility.

 An interesting insight into this mechanism is that the STN's alignment may be more robust on spectrally complex samples; spectra that are challenging for a standard classifier due to numerous peaks can provide the STN with more features to accurately determine a global shift, potentially improving performance on otherwise difficult-to-interpret data.

It is important to note that the STN developed in this work is restricted to correcting for uniform shifts, which is the expected form of variability caused by surface charging. However, non-uniform shifts can also occur, particularly in inhomogeneous samples where different parts of the surface are not at the same electrical potential (a phenomenon known as differential charging)\cite{XPSguide}. The non-uniform shifts that this introduces represent a more complex form of variance that our current STN model is not designed to handle.

\section{Future Work}

To address this limitation, future work could explore more complex transformations within the STN framework. The current model learns a single parameter for 1D translation (the shift). A promising extension would be to upgrade this to a full 1D affine transformation, which incorporates a second learnable parameter for scaling (i.e., a `squeeze' or `stretch'). Such a model could learn to not only align the spectra but also correct for uniform peak broadening by compressing or expanding the energy axis, thereby addressing a key artifact of differential charging.

The immediate next steps for this work will be to extend the current classification models to full quantification. This would involve transitioning from predicting binary labels (presence/absence) to predicting continuous compositional values for each chemical environment. Such a regression-based approach would provide a more complete and practically useful analysis, moving from simple environment detection based comparison framework to determining precise surface stoichiometry.

Furthermore, future models would benefit from adopting a more granular labelling system that defines distinct atomic environments rather than broad functional group (FG) proxies. This atom-centric approach would create a more universal format, enhancing the model's transferability to non-polymeric systems. Crucially, it would also provide a more rigorous validation of the STN's alignment capabilities, as the current FG labels may inadvertently simplify the task by implicitly linking related peaks (e.g., the C 1s and O 1s in an alcohol). An atomic-level labelling system would force the model to learn these correlations from the spectral data alone.

In addition, a more advanced avenue involves developing a flexible encoding scheme to move beyond a rigid, closed-set of labels, enabling the model to make out-of-distribution (OOD) predictions for environments not seen during training. Instead of just matching patterns, the model would need to develop a form of informed intuition, for instance, by first identifying an element and then using the magnitude of the chemical shift to infer properties like oxidation state or nearest-neighbour interactions. This would significantly enhance the model's generalizability and represent a step towards more human-like spectral interpretation, ultimately opening up its application to the vast combinatorial space of all possible chemical environments, including those that may be rare or non-physical (even if the majority of which are rare or non-physical).

While the STN-NN architecture demonstrates robust shift-invariance for complex organic polymers, it is important to acknowledge its inherent limitations. The STN thrives on polymer datasets by leveraging the stoichiometric coupling of interdependent environments (e.g., corresponding C 1s and O 1s signals) as internal reference frames for global alignment. However, for single-element systems with multiple possible oxidation states, such as unreferenced transition metals, this global alignment faces an inherent ambiguity. Without a secondary referencing element, an electrostatic charging shift can make one oxidation state positionally indistinguishable from another i.e., an Fe(II) 1s peak subject to an electrostatic shift from surface charging could appear at the exact same BE as Fe(III).

One method of resolving this ambiguity requires evaluating peak shape and multiplet splitting through shape-sensitive architectures, such as convolutional layers. A key advantage of the STN is its modular flexibility; although this study paired it with a plain MLP backbone to isolate alignment performance, the STN can also integrate upstream of more complex architectures, including Convolutional Neural Networks (CNNs), to better adapt to varied, shape-dependent datasets like transition metals.

\section{\label{sec:level6}Conclusion}
\setlength{\parskip}{0pt}

In summary, this novel implementation of a Spatial Transformer Network (STN) provides a promising solution to the challenge of variable uniform shifts induced by surface charging. As a lightweight and efficient architecture, it demonstrates a marked improvement in robustness over conventional models, such as MLPs and CNNs, for this specific task. This approach provides a promising research avenue for developing more quantitative models (rather than the simple classification models implemented here) and testing them on larger datasets. More broadly, the STN's inherent ability to account for translational variance makes it an invaluable tool for enhancing the reliability of automated analysis across a wide range of spectroscopic techniques, representing a key step towards fully autonomous laboratories.

\section{Code and Data Availability}

The code, including trained models and model architectures used to generate these results, is available at: https://github.com/Issa-Saddiq/AutoXPS. The training data used in this study were obtained under a commercial license (with express permission from the authors) and are subject to access restrictions. As such, they cannot be shared publicly. Researchers interested in accessing similar data may obtain a license directly from Scienta300 ESCA Polymer Database under equivalent terms.

\begin{acknowledgments}

This work was supported by EPSRC project EP/Y000552/1 and EP/Y014405/1. KTB acknowledges a startup fund from UCL which supported IS. We would like to thank the UCL MAPS summer internship scheme/ Ivan Parkin, The Dean of MAPS at UCL for funding the placement of Yuxin Fan for this project. MAI, PGP and DM acknowledge EPSRC funding for the National Facility for XPS ``HarwellXPS'' (EP/Y023587/1, EP/Y023609/1, EP/Y023536/1, EP/Y023552/1 and EP/Y023544/1).

\end{acknowledgments}


\bibliography{apssamp}

\newpage

\appendix*
\onecolumngrid
\section{Model Architectures}
\label{sec:architectures} % Changed label

\begin{table}[h] % Use h for here, t for top, b for bottom, p for page of its own
    \centering
    \caption{Multilayer Perceptron (MLP) Classifier Architecture Summary. $I$=input\_features (6601), $O$=output\_features (40).}
    \label{tab:classifier_arch}
    \begin{tabular}{@{}lll@{}} % @{} removes padding at table edges
        \toprule
        Component & Layer Type & Configuration / Activation \\
        \midrule
        % --- Classifier Layers ---
        \multirow{9}{*}{Classifier} & Input & Features: $I=6601$ \\
         & Linear & Units: 256 \\
         & BatchNorm1d & Features: 256 \\
         & Activation & LeakyReLU(0.01) \\
         & Dropout & Rate: 0.1 \\
         & Linear & Units: 128 \\
         & BatchNorm1d & Features: 128 \\
         & Activation & LeakyReLU(0.01) \\
         & Dropout & Rate: 0.1 \\
         & Linear (Output Logits) & Units: $O=40$ \\
        \bottomrule
    \end{tabular}
\end{table}

\begin{table}[h!]% Use h for here, t for top, b for bottom, p for page of its own
    \centering
    \caption{Convolutional Neural Network (CNN) Classifier Architecture Summary. $I$=input\_features (6601), $O$=output\_features (40).}
    \label{tab:conv_classifier_arch}
    \begin{tabular}{@{}lll@{}} % @{} removes padding at table edges
        \toprule
        Component & Layer Type & Configuration / Activation \\
        \midrule
        % --- Classifier Layers ---
        \multirow{14}{*}{Classifier} & Input & Features: $I=6601$ \\
         & Reshape & Add Channel dim: (B, 1, $I$) \\ % B = batch size
         & Conv1d & Filters: 16, Kernel: 35, Pad: 17 ('same') \\
         & BatchNorm1d & Features: 16 (applied on channels) \\
         & Activation & LeakyReLU(0.01) \\
         & Flatten & Output Dim: $16 \times I$ \\
         & Linear & Units: 256 \\
         & BatchNorm1d & Features: 256 \\
         & Activation & LeakyReLU(0.01) \\
         & Dropout & Rate: 0.1 \\
         & Linear & Units: 128 \\
         & BatchNorm1d & Features: 128 \\
         & Activation & LeakyReLU(0.01) \\
         & Dropout & Rate: 0.1 \\
         & Linear (Output Logits) & Units: $O=40$ \\
        \bottomrule
    \end{tabular}
\end{table}

\begin{table}[h!]  % Use h for here, t for top, b for bottom, p for page of its own
    \centering
    \caption{Spatial Transformer Network (STN) Classifier Architecture Summary. $I$=input\_features (6601), $O$=output\_features (40).}
    \label{tab:stn_classifier_arch}
    \begin{tabular}{@{}lll@{}} % @{} removes padding at table edges
        \toprule
        Component & Layer Type & Configuration / Activation \\
        \midrule
        % --- STN Alignment Module ---
        \multirow{6}{*}{STN Alignment} & Input & Features: $I=6601$ \\
         & \textit{Localization Network:} & \\
         & \hspace{1em} Linear & Units: 64, Activation: ReLU \\ % Indented to show sub-network
         & \hspace{1em} Linear & Units: 32, Activation: ReLU \\
         & \hspace{1em} Linear & Units: 1 (Output: translation parameter) \\
         & \textit{Spatial Transform} & Affine grid generation \& grid sampling (1D) \\
        \midrule
        % --- Classifier Module ---
        \multirow{9}{*}{Classifier} & Input & Aligned Features: $I=6601$ (from STN) \\
         & Linear & Units: 256 \\
         & BatchNorm1d & Features: 256 \\
         & Activation & LeakyReLU(0.01) \\
         & Dropout & Rate: 0.1 \\
         & Linear & Units: 128 \\
         & BatchNorm1d & Features: 128 \\
         & Activation & LeakyReLU(0.01) \\
         & Dropout & Rate: 0.1 \\
         & Linear (Output Logits) & Units: $O=40$ \\
        \bottomrule
    \end{tabular}
\end{table}

\clearpage
\onecolumngrid % Switches to single column for easier reading of large SI tables/figures

\section*{Supplementary Information}

% Reset counters
\setcounter{section}{0}
\setcounter{equation}{0}
\setcounter{figure}{0}
\setcounter{table}{0}
\setcounter{page}{1}

% Add 'S' prefix to identifiers
\renewcommand{\thesection}{S\arabic{section}}
\renewcommand{\theequation}{S\arabic{equation}}
\renewcommand{\thefigure}{S\arabic{figure}}
\renewcommand{\thetable}{S\arabic{table}}

\subsection{Synthetic Data Generation}

\begin{figure}[H] % Placement suggestion: here, top, bottom, page
    \centering % Center the entire figure content horizontally

    % --- Top row with two images ---
    \begin{subfigure}[c]{0.48\linewidth} % [c] vertically centers the subfigures
        \centering
        \includegraphics[width=\textwidth]{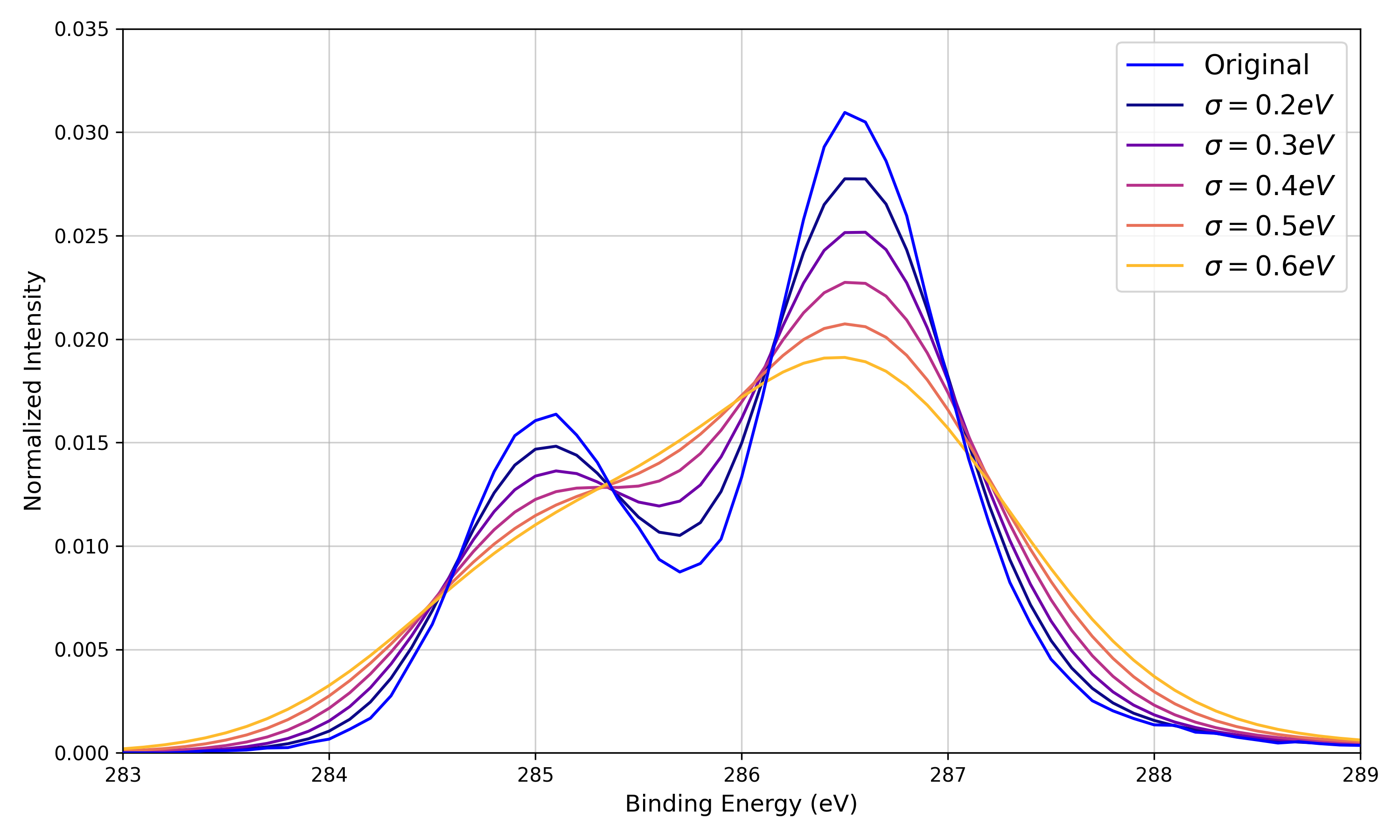} 
    \end{subfigure}
    \hfill % This automatically pushes the two images apart evenly
    \begin{subfigure}[c]{0.48\linewidth}
        \centering
        \includegraphics[width=\textwidth]{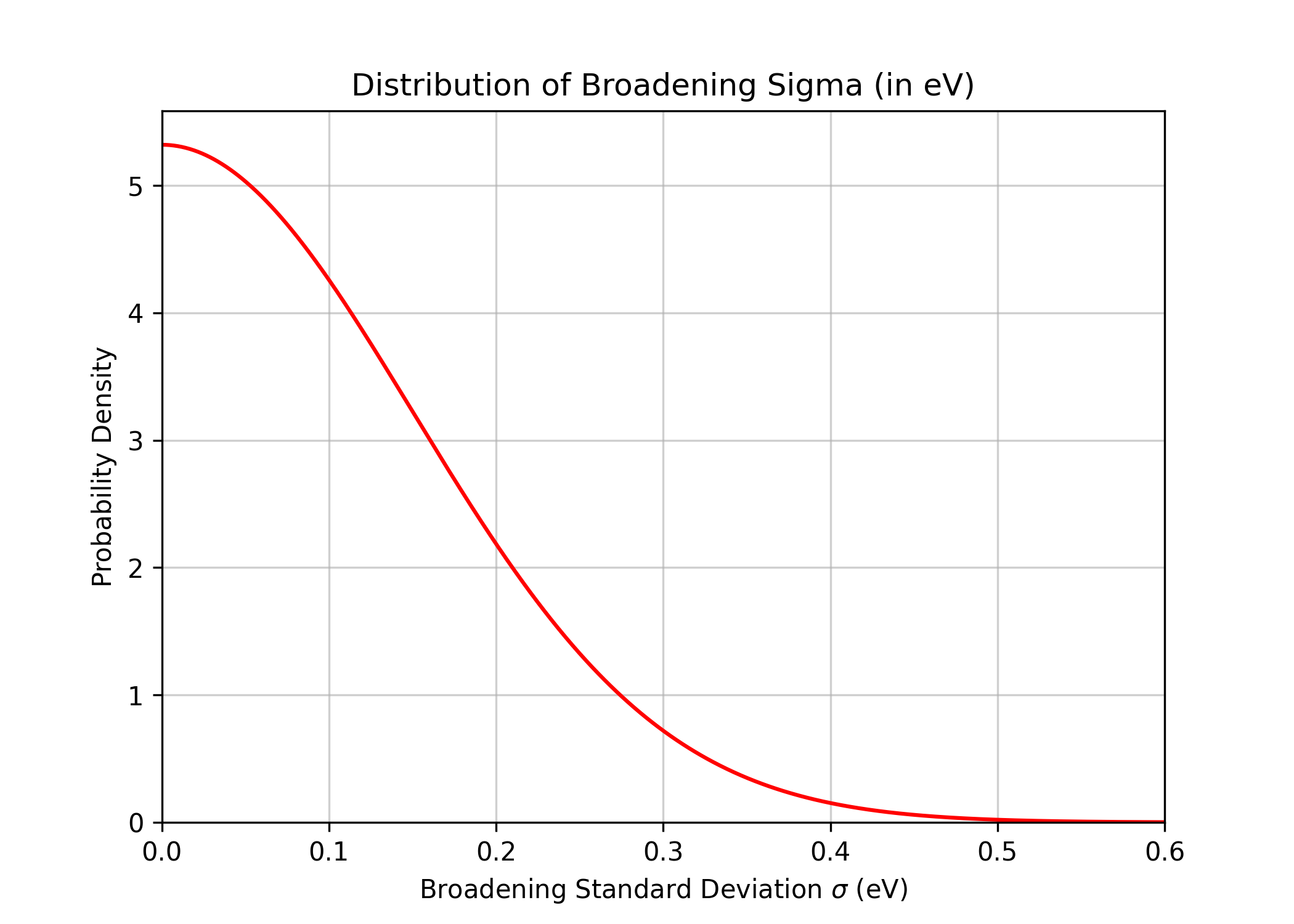}
    \end{subfigure}

    % --- Overall Caption and Label for the entire figure ---
    \caption{The increasing effects of Gaussian broadening with increasing sigma (std. dev. of Gaussian kernel), and the chosen half-normal distribution set up to enforce random broadening to differing degrees.}
    \label{fig:broadening}
\end{figure}

\subsection{Extended Performance Metrics}

\begin{figure}[ht] % use * for spanning both columns
    \centering
    % --- First subfigure ---
    \begin{subfigure}[t]{0.49\textwidth}
        \centering
        \includegraphics[width=\linewidth]{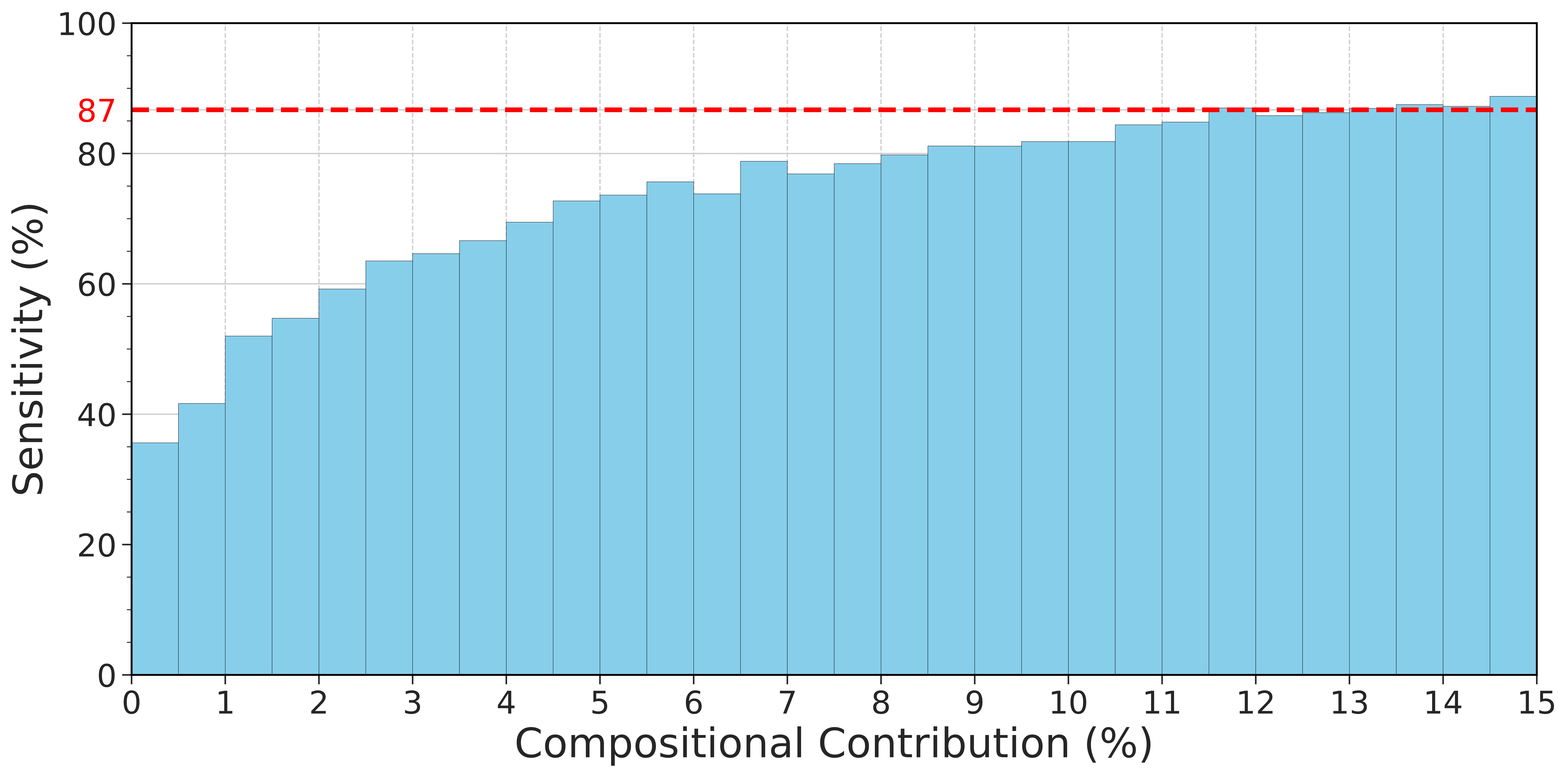}
        \caption{MLP model}
        \label{fig:MLP_detection_limits}
    \end{subfigure}
    \hfill
    % --- Second subfigure ---
    \begin{subfigure}[t]{0.49\textwidth}
        \centering
        \includegraphics[width=\linewidth]{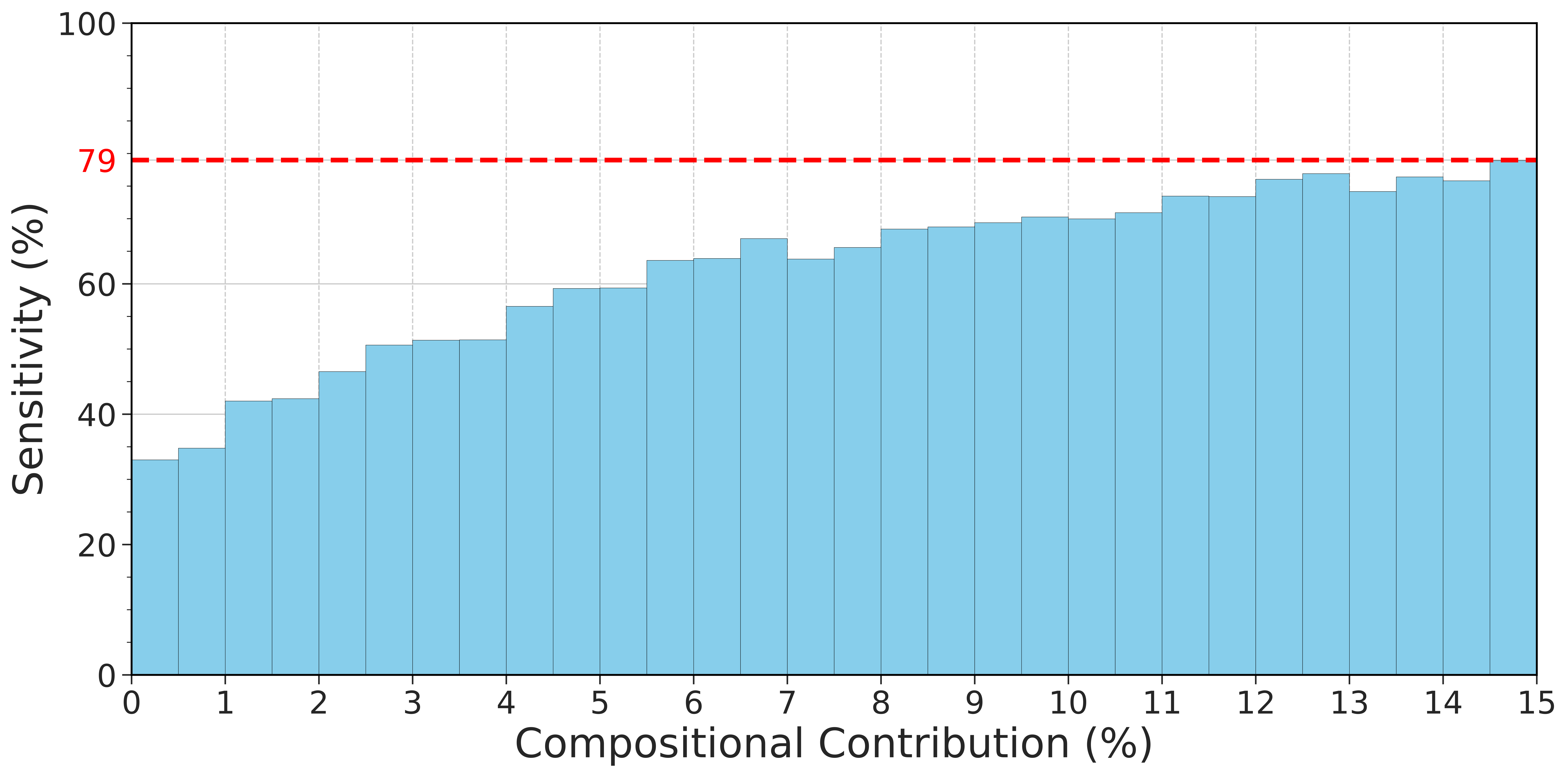}
        \caption{CNN model}
        \label{fig:CNN_detection_limits}
    \end{subfigure}

    \caption{Histograms showing the sensitivity of (a) the MLP model and (b) the CNN model as a function of compositional contribution (\%) across all functional group predictions, evaluated on spectra with a maximum 3.0 eV random shift applied. The horizontal red line marks the overall average sensitivity across all composition bins.}
    \label{fig:detection_limits_comparison}
\end{figure}

\begin{table}[H] % Use table* to span two columns, or just [htbp] for single column
    \centering
    \caption{Sensitivities (\%) of all FG classes, ordered from lowest-to-highest for the MLP model, and compared with the corresponding sensitivities for the CNN and STN-NN models. All models were trained and tested on XPS spectra with a 3eV maximum shift.}
    \label{tab:full_fg_sensitivity}
    \begin{tabular}{@{}lrrr@{}} % l for left-aligned text, r for right-aligned numbers
        \toprule
        & \multicolumn{3}{c}{Sensitivity (\%)} \\
        \cmidrule(l){2-4}
        \textbf{Functional Group} & \textbf{MLP} & \textbf{CNN} & \textbf{STN-NN} \\
        \midrule
        Epoxide              &  5.9 &  0.0 & 63.1 \\
        Anhydride            & 37.5 &  5.6 & 79.2 \\
        Alcohol (Aliphatic)  & 43.5 & 24.3 & 78.0 \\
        Naphthalene          & 46.6 &  2.9 & 88.0 \\
        Carboxylic Acid      & 52.7 & 27.7 & 84.4 \\
        Alkene               & 56.1 & 29.1 & 83.8 \\
        Ketone (Aromatic)    & 57.0 & 27.2 & 88.2 \\
        Ester (Aromatic)     & 66.3 & 46.8 & 92.2 \\
        Siloxane             & 68.4 & 46.6 & 94.6 \\
        Ketone (Aliphatic)   & 69.8 & 53.7 & 91.9 \\
        Alcohol (Aromatic)   & 70.7 & 41.7 & 94.7 \\
        Ether (Aromatic)     & 77.7 & 60.6 & 94.8 \\
        Imide                & 80.8 & 61.4 & 94.8 \\
        Ether (Aliphatic)    & 83.1 & 76.9 & 93.5 \\
        Nitrile              & 85.4 & 72.4 & 95.7 \\
        Benzene Ring         & 87.4 & 83.0 & 96.0 \\
        Ester (Aliphatic)    & 87.6 & 84.9 & 93.9 \\
        Amine                & 89.9 & 75.0 & 95.9 \\
        Sulfonate            & 90.4 & 77.7 & 97.4 \\
        Pyridine             & 90.5 & 71.6 & 96.9 \\
        Phenoxy              & 91.1 & 77.2 & 97.9 \\
        Phosphazene          & 91.1 & 77.8 & 97.9 \\
        Sulfide              & 92.5 & 75.9 & 97.1 \\
        Nitro (Aromatic)     & 93.5 & 82.5 & 97.7 \\
        Amide                & 94.2 & 88.8 & 97.5 \\
        Sulfone              & 94.6 & 88.1 & 98.0 \\
        Nitrate              & 96.3 & 87.8 & 99.2 \\
        Alkyl Halide (Cl)    & 96.5 & 92.1 & 98.6 \\
        Chlorobenzene        & 96.9 & 87.6 & 98.7 \\
        Bromobenzene         & 97.5 & 91.2 & 97.9 \\
        Alkyl Halide (F)     & 98.6 & 96.6 & 99.5 \\
        Alkane               & 99.7 &100.0 & 99.2 \\
        \bottomrule
    \end{tabular}
\end{table}

\subsection{Training Details and Convergence}

All models were developed using the PyTorch framework. To ensure a robust evaluation, the synthetic dataset was divided into an 80\% training set and a 20\% test set. Training was performed on a Tesla T4 GPU, with each model requiring approximately 15 minutes (30 minutes for plain MLP) to reach convergence over 50 epochs. Final model performance was monitored via Binary Cross Entropy loss, with the exact match accuracy used as the primary evaluation metric.

\begin{figure}[H]
    % Top Row: MLP and CNN side-by-side
    \begin{subfigure}[t]{0.48\textwidth}
        \centering
        \includegraphics[width=\textwidth]{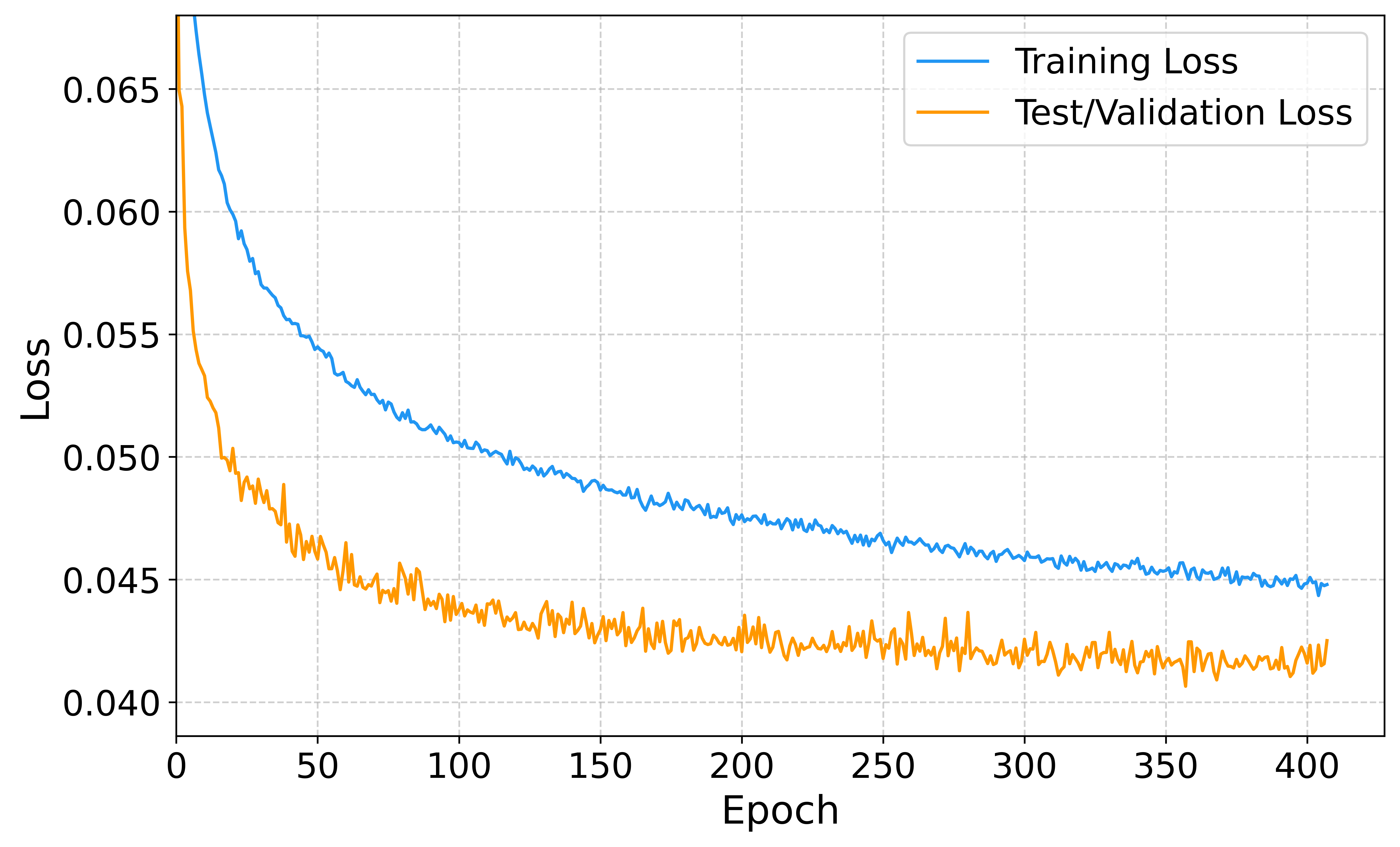}
        \caption{MLP model}
        \label{fig:mlp_curve}
    \end{subfigure}
    \hfill
    \begin{subfigure}[t]{0.48\textwidth}
        \centering
        \includegraphics[width=\textwidth]{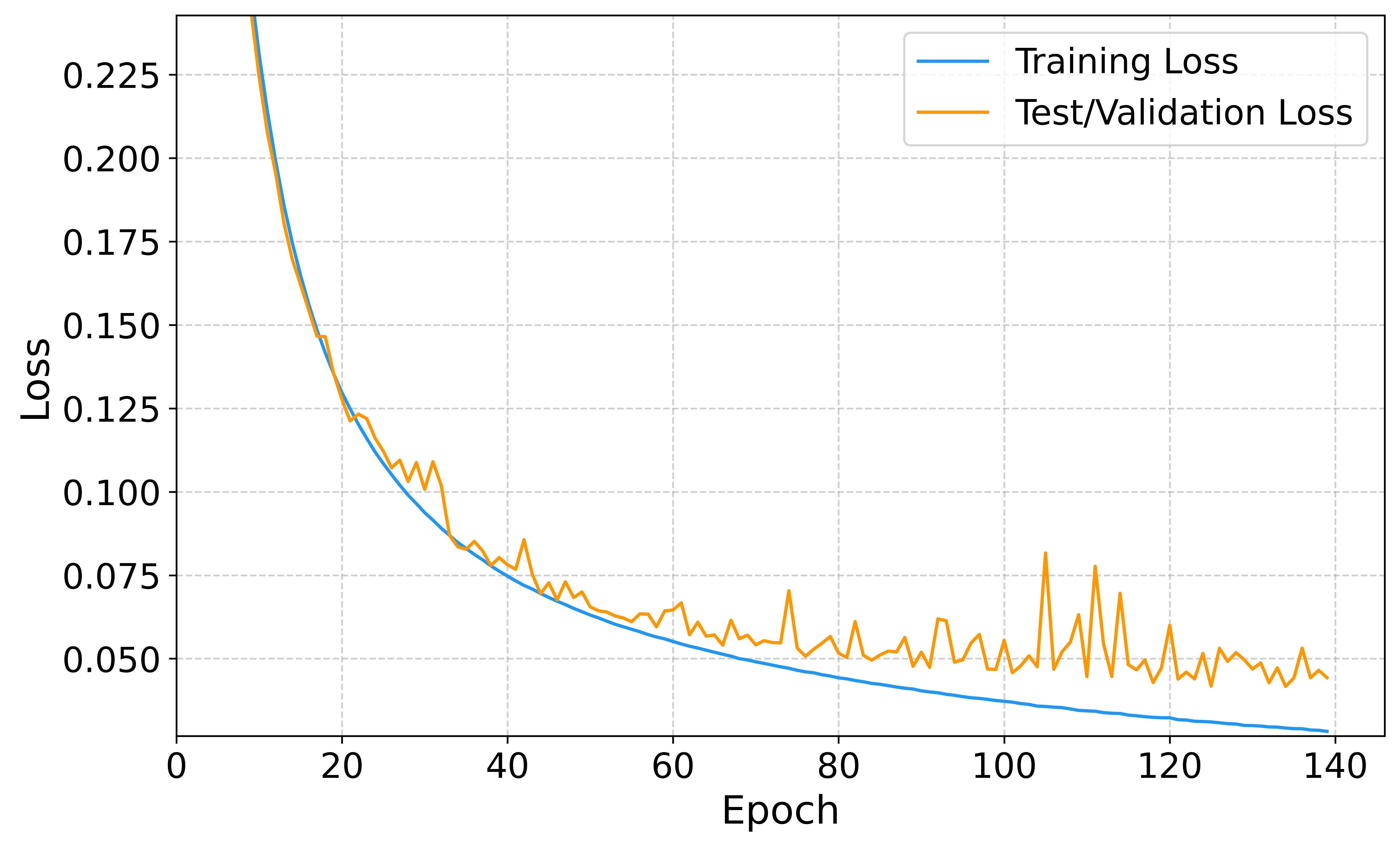}
        \caption{CNN model}
        \label{fig:cnn_curve}
    \end{subfigure}
    
    \vspace{2em} % Adds a vertical gap before the bottom image
    
    % Bottom Row: STN-NN aligned to the left
    \begin{subfigure}[t]{0.48\textwidth}
        \centering
        \includegraphics[width=\textwidth]{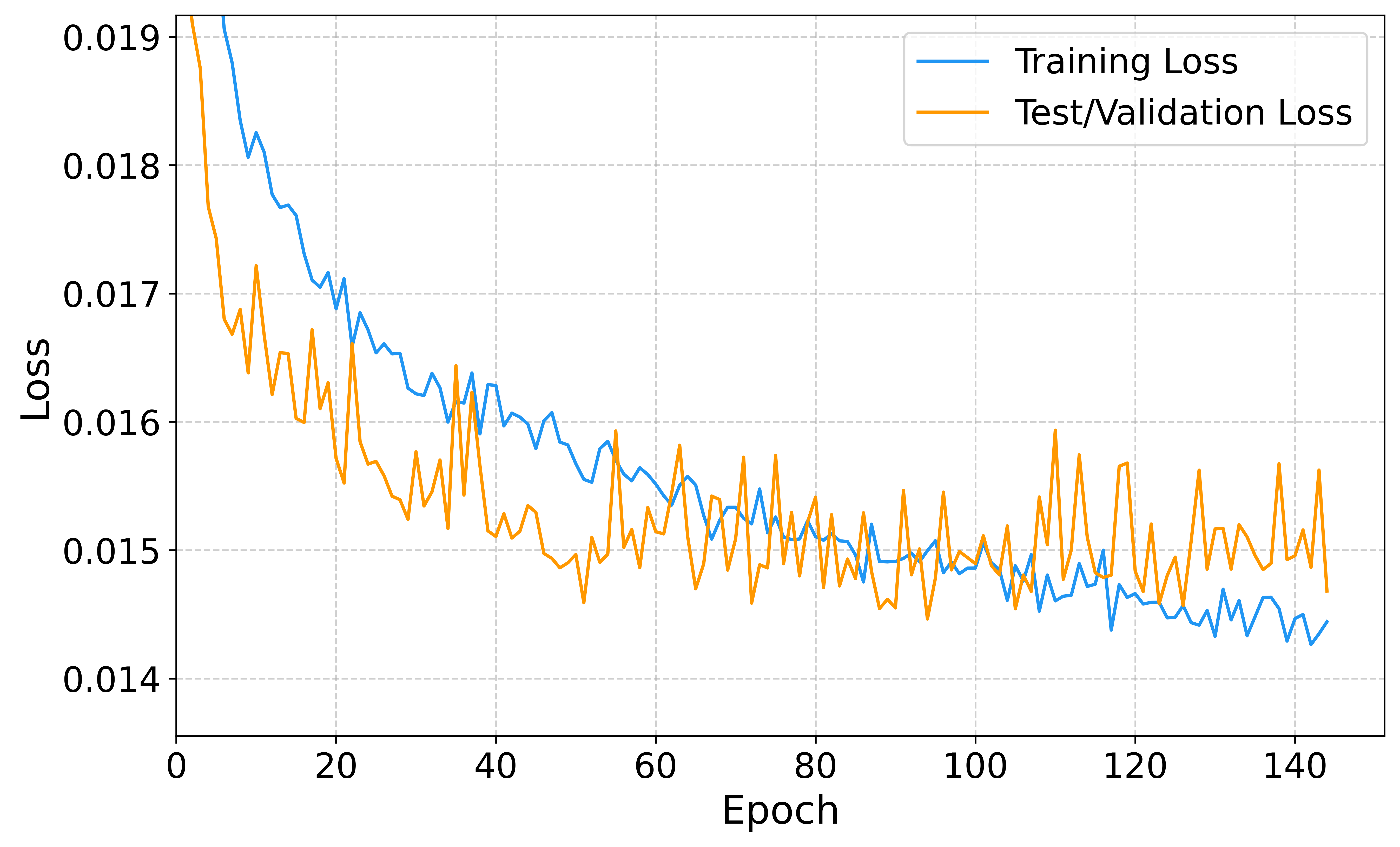}
        \caption{STN-NN model}
        \label{fig:stn_curve}
    \end{subfigure}
    \hfill % This empty hfill pushes the bottom subfigure to the left
    
    % Single unified caption for all three subplots
    \caption{Training and validation curves for the (a) MLP, (b) CNN, and (c) STN-NN models trained on spectra with a maximum 3.0 eV random shift applied.}
    \label{fig:all_training_curves}
\end{figure}

\end{document}